\documentclass[natbib]{sn-jnl}
\usepackage{natbib}

\usepackage{graphics}%
\usepackage{multirow}%
\usepackage{amsmath,amssymb,amsfonts}%
\usepackage{amsthm}%
\usepackage{mathrsfs}%
\usepackage[title]{appendix}%
\usepackage[dvipsames,table]{xcolor}%

\definecolor{myred}{RGB}{158, 15, 20}

\definecolor{blue}{RGB}{20, 96, 219}

\definecolor{myblue}{RGB}{32, 80, 158}

\usepackage{scalerel}

\usepackage{textcomp}%
\usepackage{manyfoot}%
\usepackage{booktabs}%
\usepackage{algorithm}%
\usepackage{algorithmicx}%
\usepackage{algpseudocode}%
\usepackage{listings}%
\usepackage{apacdoc}

\usepackage{adjustbox}
\usepackage{array}
\usepackage{arydshln}
\usepackage{boldline}
\usepackage{colortbl}
\usepackage{tikz}
\usetikzlibrary{arrows.meta,
	calc, chains,
	decorations.pathreplacing,%
	calligraphy,
	shapes,
	shapes.geometric, arrows, matrix, positioning, fit, calc}
\definecolor{blue1}{RGB}{84,141,212}
\definecolor{blue2}{RGB}{142,180,227}
\definecolor{yellow1}{RGB}{255,229,153}
\definecolor{orange1}{RGB}{255,153,0}
\definecolor{gray1}{RGB}{127,127,127}
\definecolor{gray2}{RGB}{217,217,217}

\usepackage{bm}
\usepackage{dsfont}
\usepackage{pdflscape}
\usepackage{ltxtable, tabularx}
\usepackage{geometry}

\newcounter{alphasect}
\def\alphainsection{0}

\let\oldsection=\section
\def\section{%
	\ifnum\alphainsection=1%
	\addtocounter{alphasect}{1}
	\fi%
	\oldsection}%

\renewcommand\thesection{%
	\ifnum\alphainsection=1%
	\Alph{alphasect}
	\else%
	\arabic{section}
	\fi%
}%

\makeatletter
\renewcommand*\l@figure{\@dottedtocline{1}{1em}{2em}}
\let\l@table\l@figure
\makeatother

\usepackage{bbm}
\usepackage{float}

\usepackage{subcaption}
\usepackage{caption}
\usepackage{stackengine}
\DeclareMathOperator*{\argmax}{arg\,max}

\theoremstyle{thmstyleone}%
\theoremstyle{thmstyletwo}%

\theoremstyle{thmstylethree}%

\begin{document}

	\title[Article Title]{Linear classification methods for multivariate repeated measures data - a simulation study \\
		{\small Short title: Multivariate repeated measures data classification}}

	
		\author*[1]{\fnm{Ricarda} \sur{Graf}}\email{ricarda.graf@math.uni-augsburg.de}
		
		\author[2,3]{\fnm{Marina} \sur{Zeldovich}}\email{Marina.Zeldovich@uibk.ac.at}
		
		\author[1,4]{\fnm{Sarah} \sur{Friedrich}}\email{sarah.friedrich@math.uni-augsburg.de}

		\affil[1]{\orgdiv{Department of Mathematics}, \orgname{University of Augsburg}, \orgaddress{\street{Universit{\"a}tsstra{\ss}e 2}, \city{Augsburg}, \postcode{86159},  \country{Germany}}}
		
		\affil[2]{\orgdiv{Institute of Psychology}, \orgname{University of Innsbruck}, \orgaddress{\street{Universit{\"a}tsstra{\ss}e 5-7}, \city{Innsbruck}, \postcode{6020}, \country{Austria}}}
		\affil[3]{\orgdiv{Faculty of Psychotherapy Science}, \orgname{Sigmund Freud University Vienna}, \orgaddress{\street{Freudplatz 1}, \city{Vienna}, \postcode{1020}, \country{Austria}}}
		
		\affil[4]{\orgdiv{Centre for Advanced Analytics and Predictive Sciences (CAAPS)}, \orgname{University of Augsburg}, \orgaddress{\street{Universit{\"a}tsstra{\ss}e 2}, \city{Augsburg}, \postcode{86159}, \country{Germany}}}

	
	\abstract{
		Researchers in the behavioral and social sciences  use linear discriminant analysis (LDA) for predictions of group membership (classification) and for identifying the variables most relevant to group separation among a set of continuous correlated variables (description).  \\ 
		In these and other disciplines, longitudinal data are often collected which provide additional temporal information. Linear classification methods for repeated measures data are more sensitive to actual group differences by taking the complex correlations between time points and variables into account, but are rarely discussed in the literature. Moreover, psychometric data rarely fulfill the multivariate normality assumption.\\
		In this paper, we compare existing linear classification algorithms for nonnormally distributed multivariate repeated measures data in a simulation study based on  psychological questionnaire data comprising Likert scales. 
		 The results show that in data without any specific assumed structure and larger sample sizes, the robust alternatives  to standard repeated measures LDA may not be needed. To our knowledge, this is one of the few studies discussing repeated measures classification techniques, and the first one comparing multiple alternatives among each other.

		
	}

	
	
	\maketitle
	

	\section{Introduction}\label{sec1}
	
	In psychology and the social sciences, discriminant analysis (DA) is traditionally applied to classification tasks in data with continuous variables since its invention by Fisher (\citeyear{Fisher1936}). Based on estimates of group means and the pooled covariance matrix, a classification rule is obtained or relative variable weights can be computed, respectively.
	Its importance for the behavioral sciences has often been emphasized in reviews, tutorials and textbooks \citep{Boedeker2019, Sherry2006, Field2018, Huberty2006, Fletcher1978, Betz1987, Garrett1943}. It has  been applied to a large number of problems in experimental and applied psychology for class prediction as well as description \citep{Rogge1999, Langlois2000, OBrien2009, Kumpulainen2018, Shinba2021, Stoyanov2022, Aggarwala2022}.\\
	In contrast to multivariate data measured at a single time point, longitudinal data provide additional information about temporal changes, wherefore they are collected in various disciplines, including psychology and the social sciences \citep{Jensen2021, Banks2021, Lanahan2019}. Despite these potential applications for repeated measures DA or alternative linear classification techniques, textbooks discussing DA do not mention respective repeated measures approaches \citep{Lix2010}. \\
	To complicate matters further, many classification approaches for continuous multivariate repeated measures data assume multivariate normality  \citep{Roy20051, Roy20052, Tomasko1999, Gupta1986}, but this assumption is rarely fulfilled by psychological datasets and hard to verify for small sample sizes \citep{Delacre2017, Rausch2009, Beaumont2006, Neto2016}. Psychological data, especially those obtained using patient-reported instruments, are often characterized by skewness.\\
	There are only few alternative repeated measures approaches which relax or overcome the multivariate normality assumption and take the complex correlation structure between time points and variables into account. It is the aim of  this manuscript to compare these approaches in an extensive simulation study. In particular, we consider two modifications of repeated measures LDA by Brobbey et al. (\citeyear{Brobbey2021, Brobbey2022}) that are more robust to deviations from multivariate normality, and the generalization of the support vector machine classifier by Chen and Bowman (\citeyear{Chen2011}) to longitudinal data, which is a nonparametric linear classifier when used with a linear kernel. We compare these methods' performance among each other and choose more general, realistic simulation settings, including unequal sample sizes, unstructured covariance matrices, and varying correlations over time instead of assuming any specific pattern.  \\
	Brobbey (\citeyear{Brobbey2021}) compares the standard repeated measures LDA (assuming multivariate normality and homoscedasticity) 
	to its performance after preceding multivariate outlier removal based on two trimming algorithms \citep{Rousseeuw1985}. In Brobbey et al. (\citeyear{Brobbey2022}),  the performance of the standard approach is compared to its performance when based on parsimonious Kronecker product structure covariance matrix estimates (\citeyear{Brobbey2022}) from the generalized estimating equations (GEE) model \citep{Inan2015}.  The longitudinal support vector machine classifier by Chen and Bowman (\citeyear{Chen2011}) uses a weighted combination of multivariate measurements taken at several time points as input in order to represent the data structure more realistically.\\
	Thus, this paper provides a neutral comparison study which evaluates the performance of the standard repeated measures LDA, its robust and nonparametric alternatives as well as all possible combinations thereof, in linear classification problems of multivariate repeated measures data and investigate their robustness when data deviate from multivariate normality. 
	In order to mimick realistic datasets, we base simulations on unstructured means and covariance matrices estimated from psychometric reference datasets which differ in sample size, sample size ratios, class overlap, temporal variation and number of measurement occasions.  In addition to method comparisons using data simulations, we evaluate the algorithms' performance in the reference data using a nonparametric bootstrap approach which provides confidence intervals for the performance measures \citep{Wahl2016}. \\
	 The paper is organized as follows. In Section \ref{data}, we explain the general structure of Likert-type data and its analysis. Some of the literature sources mention the need for longitudinal techniques. We then discuss the characteristics of the five reference datasets, which are based on Likert-type data.  
		In Section \ref{methods}, we introduce the classification algorithms whose performance we compare, and the two approaches based on the reference data and data simulations, respectively, to compare them. In Section \ref{results}, we present and discuss the results and provide recommendations based on the findings.  Conclusions are made in Section \ref{conclusion}.

	\section{Data}\label{data}
	
	Questionnaires using Likert-type responses data are a typical example of psychological data to which LDA is applied. In Section \ref{likert_data} we describe the general data structure and how LDA for linear classification is used for validating the importance of a particular subset of variables with the aim of distinguishing two groups. Some sources explicitly mention the need for longitudinal techniques, emphasizing the need for discussing available techniques. 
	In Section \ref{ref_data}, we present the two reference datasets for which individuals completed standardized questionnaires using Likert-type responses. In order to examine the methods' performance in further relevant scenarios, we additionally considered multiple modifications of these datasets, which will also be described.   
	
	\subsection{Psychological questionnaires using Likert-type scales}\label{likert_data}
	
	In psychological and social science research, behaviour is most often assessed by self-report questionnaires using Likert scales \citep{Baumeister2010,Clark2019,Sullivan2013}. It is common practice to create pools of Likert items to form subscales which each represent an aspect of the overall construct that the questionnaire is intended to investigate. Single Likert items (i.e. questions) are not considered to sufficiently capture these aspects \citep{Rickards2012,Clark2019} and are therefore summarized into subscales by considering either the sum or average of subgroups of Likert items. The development and best practices of  constructing questionnaires using Likert-type responses is discussed in the methodological psychology literature \citep{Jebb2021}.
	Likert (\citeyear{Likert1932}) developed the typical 5- or 7-point  ordinal scale on which single items are measured, e.g. ranging from ``strongly approve'' to ``strongly disapprove''. He suggests to assign numerical values to the answer choices in the same order as they are ranked. However, he does not suggest that these ordinal values must necessarily be translated into an equidistant
	scale, and states that the same results will be obtained as long as the rank order is preserved. This  translation of an ordinal scale into a numerical scale conditional on rank preservation is considered to be legitimate elsewhere \citep{Silan2020}. So in conclusion, the distances between the numerical values are irrelevant to  the analysis \citep{Gaito1980} which complies with the ordinal measurement scale of the Likert items where distances between answer choices cannot be measured. Likert (\citeyear{Likert1932}) suggests to subsequently take the sum or mean of the transformed values, which he  assumes to be normally distributed. There is a long-standing debate about how Likert-type scales should appropriately be analysed but the prevailing opinion due to vast empirical evidence \citep{Norman2010, Carifio2007} is that survey scales as opposed to single Likert items may be treated as interval data such that means and standard deviations can be computed, and parametric methods should be applied to them  \citep{Carifio2008, Rickards2012, Sullivan2013}. \\
	Specific examples for the application of LDA to questionnaire data based on Likert-type scales are Wang et al. (\citeyear{Wang2016}), Veronese and Pepe (\citeyear{Veronese2017}),  Kristjansdottir  et al. (\citeyear{Kristjansdottir2018}), and Knowles et al. (\citeyear{Knowles2000}). In all of these studies, the authors computed Fisher discriminant function coefficients (descriptive DA) for the subscales of the considered psychological questionnaires using Likert-type responses and showed the validity of these coefficients, i.e. their discriminative ability, by subsequent linear classification (predictive DA). In particular, Wang et al. (\citeyear{Wang2016}) examine a longitudinal data set but restrict their analysis to time point one
	when applying LDA. Veronese and Pepe (\citeyear{Veronese2017}) emphasize the need to explore the dynamic relations between their chosen subscales over time and point out their restriction to cross-sectional data in their LDA as a considerable limitation. 

	\subsection{Reference datasets}\label{ref_data}
	
\begin{figure}[H]
		\renewcommand\thefigure{1}
		\begin{subfigure}{\textwidth}
			\hspace{0.05cm}		
			\includegraphics[scale = 0.214]{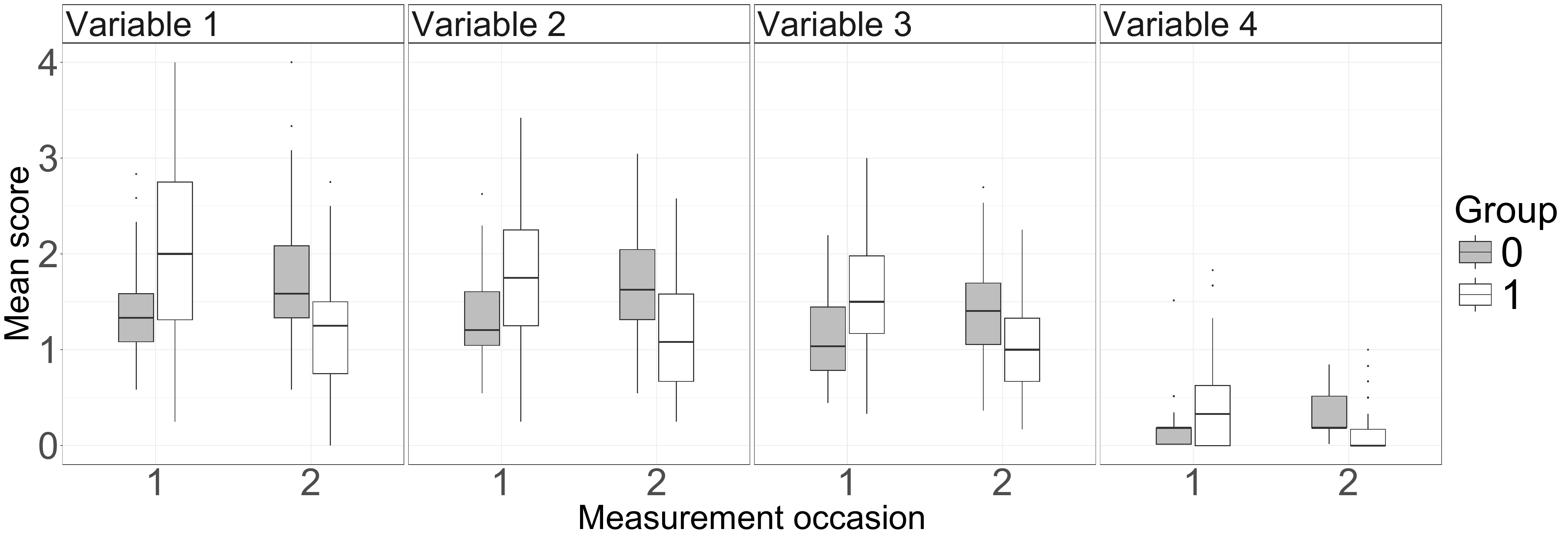} 
			\vspace{-0.57cm}
			\caption*{(c)}
		\end{subfigure}	
		\begin{subfigure}{\textwidth}
			\hspace{-0.05cm}		
			\includegraphics[scale = 0.216]{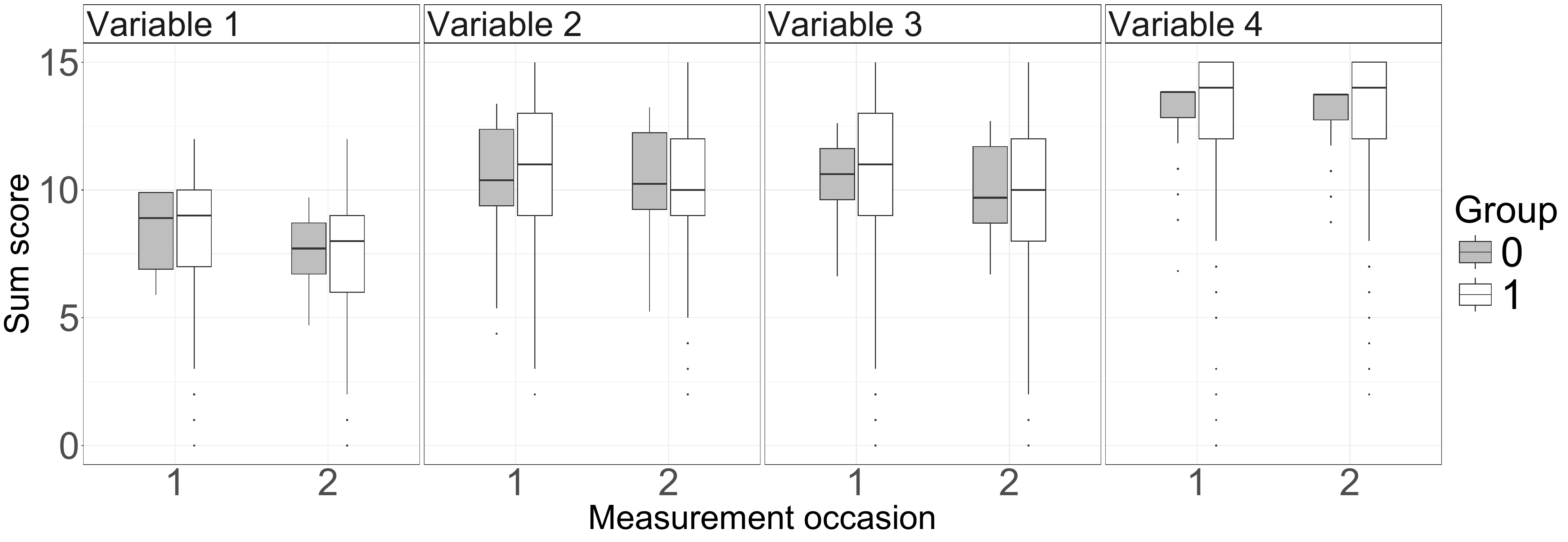} 
			\vspace{-0.57cm}
			\caption*{(d)}
		\end{subfigure}					
		\begin{subfigure}{\textwidth}
			\hspace{0.05cm}
			\includegraphics[scale = 0.22,trim={0cm, 0, 0cm, 0},clip]{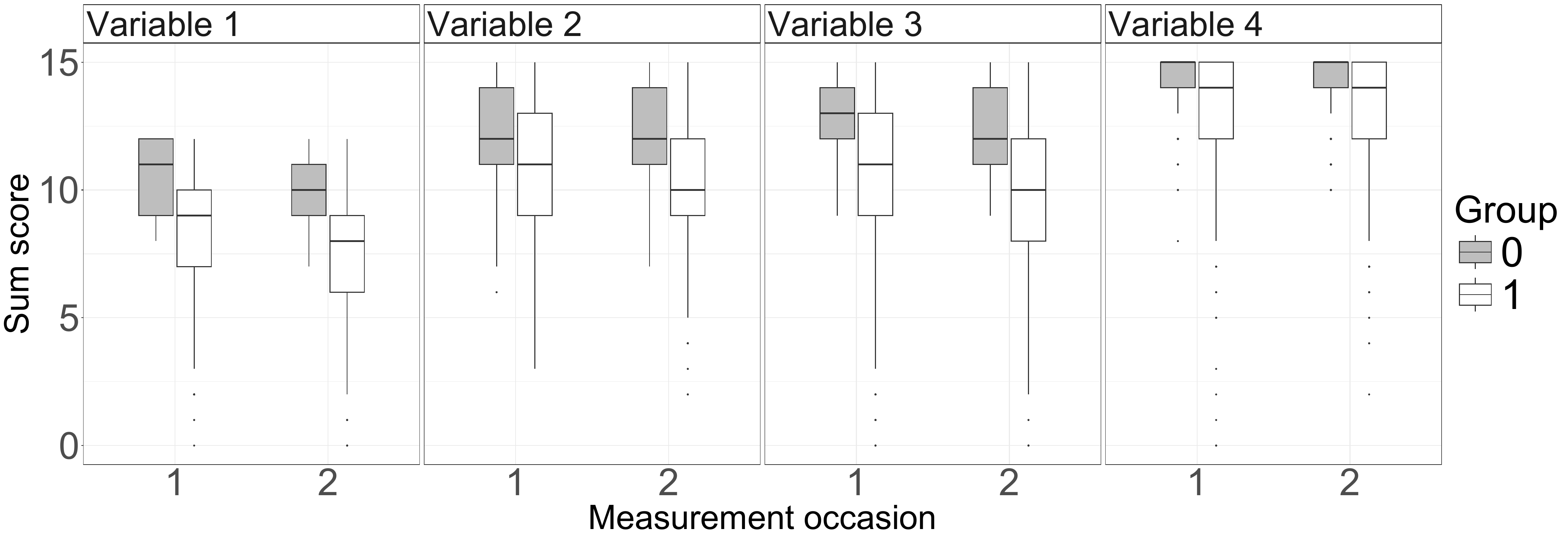}   
			\vspace{-0.57cm}
			\caption*{(e)}
			\vspace{-0.1cm}
		\end{subfigure}		
		\caption[Boxplots showing the variables' distribution in the reference datasets.]{\fontsize{10}{13}\selectfont
			{(this and previous page) Boxplots showing the variables' distribution in the reference datasets:\\
				(a) Dataset 1: CORE-OM dataset, group variable \textit{hospitalisation}  ($n_0 = 42, n_1 = 142$, non-hospitalised individuals represent group 0 and hospitalised individuals represent group 1)\\
					(b) Dataset 2: CASP-19 dataset, group variable \textit{loneliness} ($n_0 = 948, n_1 = 1682$, participants who feel less lonely represent group 0 and participants who feel more lonely represent group 1)\\
					(c) Dataset 3 (modified Dataset 1): same collapsed means, group means with opposite temporal trends\\
					(d) Dataset 4 (modified Dataset 2, time points 1 \& 2): same means, group covariance matrices differ,\\
					(e) Dataset 5 (modified Dataset 2, time points 1 \& 2): balanced class sizes by random undersampling of group 1.} 
		}
		\label{boxplots_ref}
	\end{figure}

\begin{figure}[H]
		\renewcommand\thefigure{2}	
		\begin{minipage}{.48\textwidth}
			\includegraphics[scale = 0.35]{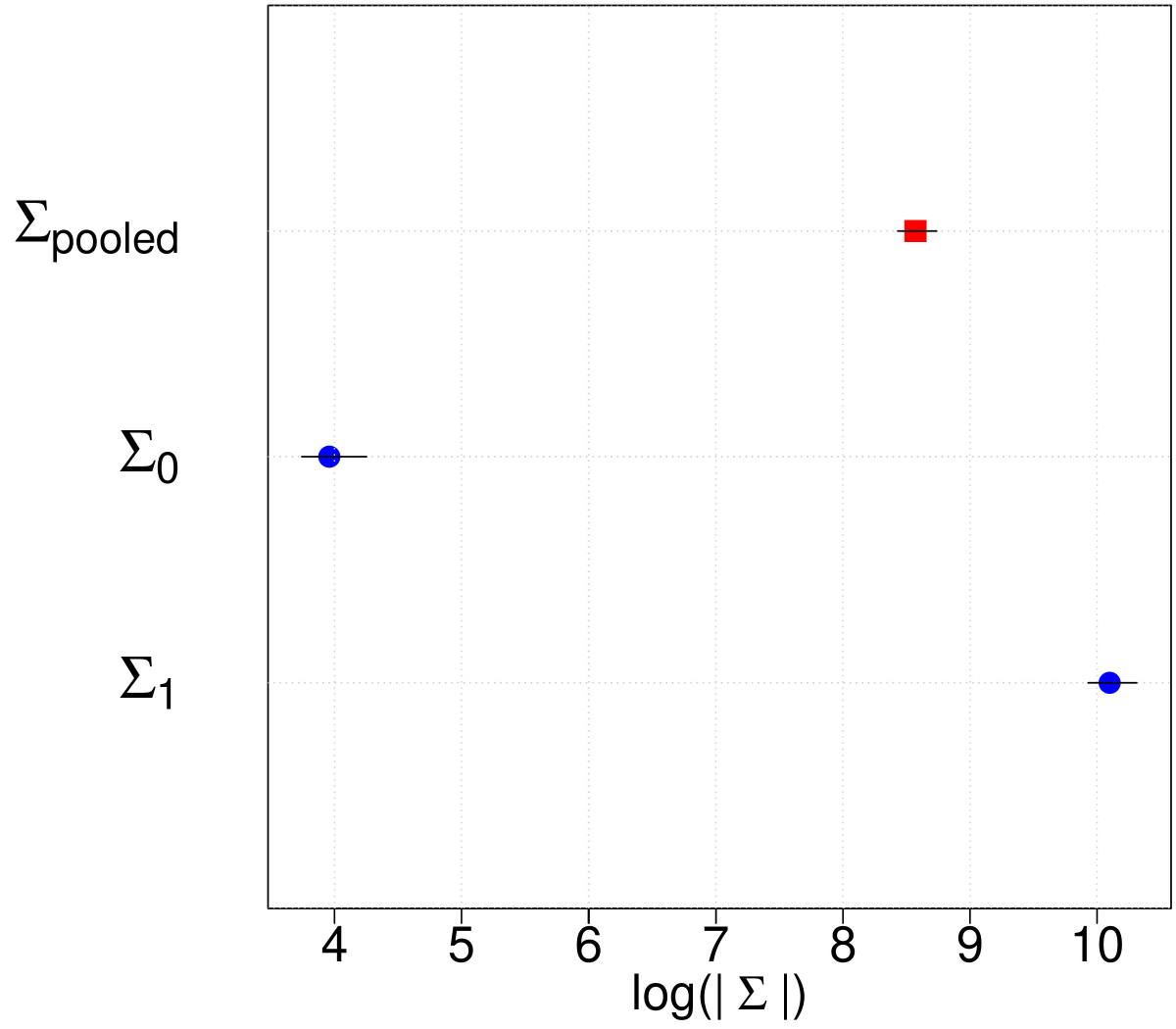}
		\end{minipage}
		\begin{minipage}{.03\textwidth}
			\hspace{0.05cm}
		\end{minipage}
		\begin{minipage}{.48\textwidth}
			\includegraphics[scale = 0.315]{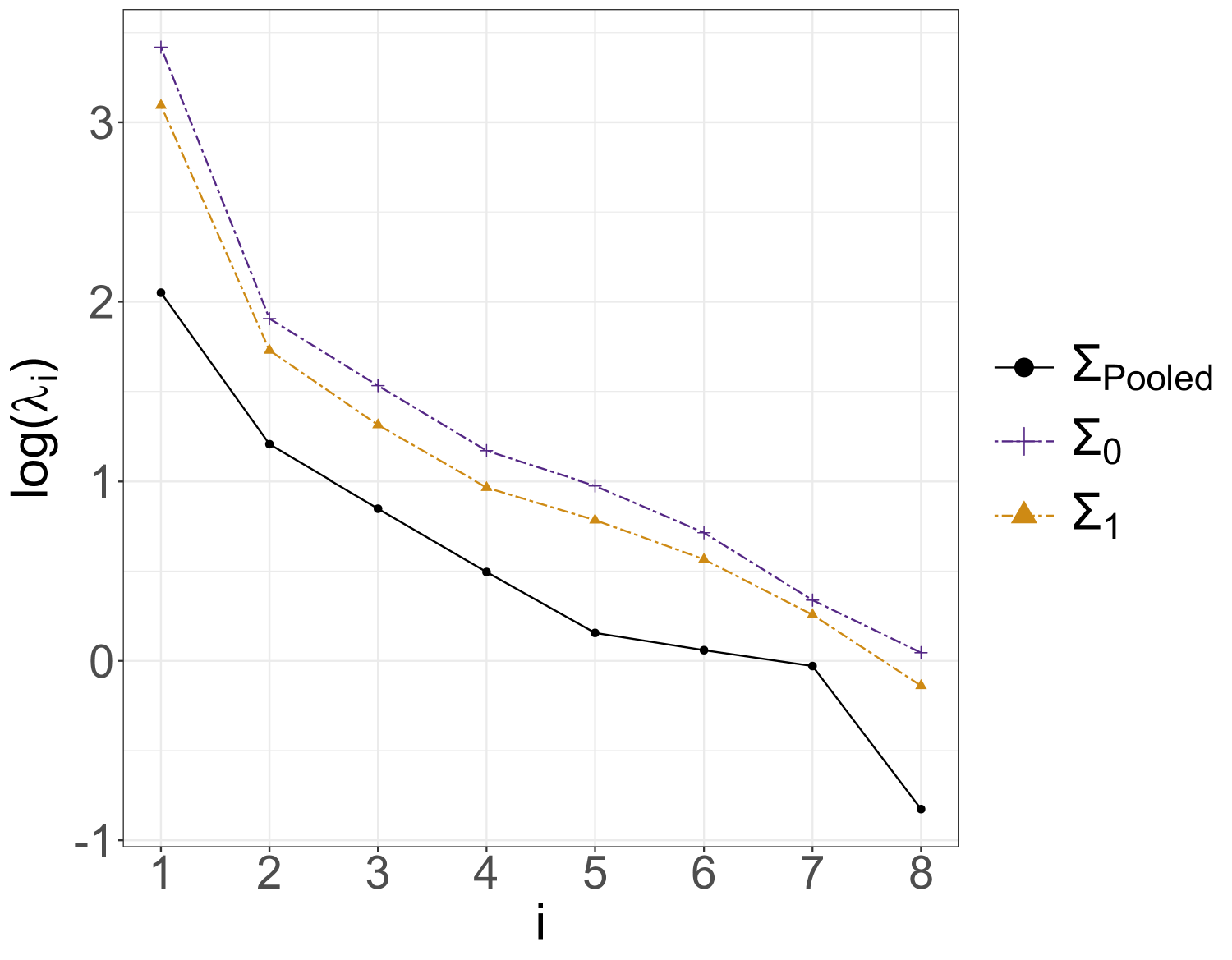}
		\end{minipage}
		\caption[Check for homoscedasticity]{\fontsize{10}{13}\selectfont
			{ Plots of the components of Box's M test for Dataset 4. Left: log determinants of covariance matrices with asymptotic 95\% confidence intervals (CI). Right: scree plots of log eigenvalues of the covariance matrices. Less overlap of CIs and higher differences between log eigenvalues, respectively, correspond to a higher degree of heterogeneity of the (group) covariance matrices. The figures indicate (significant) heterogeneity of covariance matrices.} 
		}
		\label{cov_homogeneity}
	\end{figure}

	Two datasets differing in the number of repeated measurement occasions, as well as two modifications thereof, are used as reference datasets. Each original dataset comprises measurements of four continuous predictor variables which are measured at two time points (CORE-OM dataset) and four time points (CASP-19 dataset), respectively. The binary outcome variable represents the group ($y \in \{0,1\}$). Both of these standardized psychological questionnaires consist of Likert-type questions measured on a 5-point and 4-point Likert scale, respectively. According to the developers of these questionnaires, we considered the mean score of multiple Likert items in case of the CORE-OM dataset, and the sum score in case of the CASP-19 dataset, respectively, as the basis for parameter estimation and subsequent data simulation. \\ 
	We created reference datasets from these data in order to compare the methods' performance in different (almost) realistic settings, not in order to draw any substantive conclusions about the data themselves. Datasets differ among others in sample sizes, sample size ratios, class overlap, temporal variation, and number of measurement occasions.\\
	The first dataset \citep{Zeldovich2018} is a self-report questionnaire of psychological distress abbreviated to  CORE-OM (Clinical Outcomes in Routine Evaluation-Outcome Measure) \citep{Barkham1998}. It assesses the progress of psychological or psychotherapeutic treatment using four domains (subjective well-being, problems/symptoms, life functioning, risk/harm) measured on a 5-point Likert scale (0: not at all, 1: only occasionally, 2: sometimes, 3: often, 4: most or all the time).
	Our dataset uses the binary variable hospitalisation as group variable and is denoted as ``dataset 1'' in the following. Non-hospitalised participants represent group 0 ($n_0 = 42$) and hospitalised ones group 1 ($n_1 = 142$). \\
	The second dataset is a self-report questionnaire of quality of life developed for adults aged 60 and older abbreviated to CASP-19 \citep{Hyde2003}. The dataset on CASP-19 is derived from waves 2, 3, 4, and 5 of The English Longitudinal Study of Ageing (ELSA) \citep{Banks2021}. The CASP-19 questionnaire comprises four subdomains (control, autonomy, self-realization, pleasure) measured on a 4-point Likert scale (0: often, 1: sometimes, 2: not often, 3: never; reversed scale for some items). Loneliness as one of the factors affecting quality of life \citep{Talarska2018} is chosen as the group variable. For this purpose, the sample was dichotomized at a score value of three determined from two questions related to loneliness (``Old age is a time of loneliness'', ``As I get older, I expect to become more lonely''), answered on a 5-point Likert scale (1: strongly agree, 5: strongly disagree) by the participants during wave 2.  Persons who feel less lonely represent group 0 ($n_0 = 948$) and those who feel more lonely represent group 1  ($n_1 = 1682$). Since the group differences were nevertheless marginal, we modified these data. All individuals of group 1 were included in our reference dataset, but only those individuals of group 0 were included, whose scores of the variables ``control'' and ``self-realization'' lay above their respective 0.2 percentiles. The dataset is referred to as ``dataset 2'' in the following. \\  
		Answers to questions of each subdomain in these  questionnaires using Likert-type responses are summarized in a score, where a higher mean score correspond to a higher level of distress (dataset 1), and a higher sum score indicates a better quality of life (dataset 2), respectively. Data simulations are based on these scores. Boxplots in Figure \ref{boxplots_ref}a and \ref{boxplots_ref}b show the scores' distribution in reference data 1 and 2, respectively. They indicate that on average individuals in one group usually obtain higher/lower scores compared to the other group  irrespective of the time point and variable, presumably facilitating classification in these datasets. Also, temporal variation in dataset 2  is rather modest. \\
		Therefore, we considered further scenarios beyond these two original datasets (dataset 1: CORE-OM, dataset 2: CASP-19). In addition, to test the methods under different conditions, we provided three modified versions of these datasets (dataset 3: modified CORE-OM with equal group means collapsed over time points and  group means with opposite temporal trends, dataset 4: modified CASP-19, time points 1 \& 2 only, with identical means but heterogeneous covariance matrices, dataset 5: modified CASP-19, time points 1 \& 2 only, balanced class sizes by random undersampling of group 1). Dataset 3 was modified by adding a constant specific to each variable to the data of group 0 such that collapsed means of both groups became equal in size, while maintaining the original boundaries of the measurement intervals. Then we swapped the data of the two time points for variables 1, 2, and 3 for group 0, such that means of group 0 have an upward temporal trend compared to the downward temporal trend of measurements in group 1. For dataset 4, only time points 1 and 2 are considered. We adjusted group means of group 0 per time point such that they equal those of group 1. For dataset 5, also only time points 1 and 2 are considered, and  
		a random subset of the larger group 1 equalling the sample size of group 0 was chosen in order to obtain a balanced scenario. The corresponding \texttt{R} code can be found on Figshare (see code ``availability'').\\ 
		With dataset 4, the aim was to create data which only differ in their group covariance matrices. Homogeneity of covariance matrices can be tested using the well-known Box's M test \citep{Box1949}, but its reliability suffers when the multivariate normality assumption is even only slightly violated \citep{Tiku1984}. For dataset 4, the $p$-value of the approximate $\chi^2$ test statistic of the Box's M test is $< .001$  ($\chi^2$(36) = 1789.9) but this significant test result may indicate a violation of normality instead of inequality of covariance matrices. Therefore, since the data significantly differ from multivariate normality (Table S 4), we visually assessed the covariance matrices' heterogeneity based on the components used for Box's M test, i.e. log determinants of the pooled and group covariance matrices ($\bm{\Sigma}_{\text{pooled}}, \bm{\Sigma}_0, \text{ and } \bm{\Sigma}_1$), which equal the product of their respective log eigenvalues. We use plots of log determinants with 95\% confidence intervals and plots of log eigenvalues of the covariance matrices as suggested by Friendly \& Sigal (\citeyear{Friendly2018}). From Figure \ref{cov_homogeneity} we conclude substantial heterogeneity of the group covariance matrices $\bm{\Sigma}_0$ and $\bm{\Sigma}_1$ in dataset 4. For dataset 4, simulations are based on the estimates of $\bm{\Sigma}_0$ and $\bm{\Sigma}_1$, whereas for datasets 1-3, and 5 they are based on the estimate of $\bm{\Sigma}_{\text{pooled}}$ such that the LDA assumption of homogeneous covariance matrices holds. Figure S 1 shows the plots for inspecting heterogeneity of covariance matrices for the other reference datasets as well. The assumption is not fulfilled in any of the datasets. Boxplots in Figure \ref{boxplots_ref}c-e show the scores' distribution in reference data 3-5. \\
		We chose reference datasets with moderate temporal and cross-sectional correlations. Correlation matrices are shown in Table S 1a - S 1e.  In this case, analyzing the data separately per time point or focussing on measurements of single variables over multiple time points, respectively, would ignore these correlations and yield less reliable results if, in fact, affiliation to one of the groups is affected by multiple correlated variables and/or time points (e.g. \citeauthor{Gnanadesikan1984}, \citeyear{Gnanadesikan1984}).

	\begin{table}[htb]
		\centering
		\small	
		\caption[]{\small{Some properties of the reference datasets and the corresponding simulation scenarios considered in the simulation study.
			}\\
			{  \footnotesize  Abbreviations: $\bm{\Sigma}_{\text{pooled}}$: pooled covariance matrix, $\bm{\Sigma}_0$: covariance matrix group 0, $\bm{\Sigma}_1$: covariance matrix group 1.}} 
		\begin{tabular*}{\textwidth}{@{}p{7em}p{4.5em}p{3em}p{5em}p{6em}p{12.1em}@{}} \toprule		
			&  \parbox{1.65cm}{\# Variables} &  \parbox{1.1cm}{\# Time  points}    & \parbox{1.8cm}{Sample sizes}  & \parbox{2.1cm}{Covariance matrix used for simulations}  & \parbox{4cm}{Description of simulation scenario} \\ \midrule
			${\textit{Dataset 1}}$  & 4  &	2		& \parbox{1.8cm}{$n_0 = 42\\ n_1 = 142$}	& $\bm{\Sigma}_{\text{pooled}}$ &  \parbox{4cm}{unbalanced sample sizes, homogeneous covariance matrices,
				same temporal trends of group means}         \\  \arrayrulecolor{gray!50}\cmidrule[0.05pt]{1-6}
			${\textit{Dataset 2}}$  & 4  &  4       & \parbox{1.8cm}{$n_0 = 948\\ n_1 = 1682$} & $\bm{\Sigma}_{\text{pooled}}$ & \parbox{4cm}{unbalanced sample sizes, homogeneous covariance matrices,
				same temporal trends of group means} \\ \arrayrulecolor{gray!50}\cmidrule[0.05pt]{1-6}
			${\textit{Dataset 3}}$  & 4  &	2		& \parbox{1.8cm}{$n_0 = 42\\ n_1 = 142$}	& $\bm{\Sigma}_{\text{pooled}}$  &  \parbox{4cm}{unbalanced sample sizes, homogeneous covariance matrices,
				same group means collapsed over time, opposite temporal trends} \\  \arrayrulecolor{gray!50}\cmidrule[0.05pt]{1-6}
			${\textit{Dataset 4}}$  & 4  &	2		& \parbox{1.8cm}{$n_0 = 948\\ n_1 = 1682$}	& $\bm{\Sigma}_0, \bm{\Sigma}_1$ & \parbox{4cm}{unbalanced sample sizes, heterogeneous covariance matrices,
				same group means}   \\  \arrayrulecolor{gray!50}\cmidrule[0.05pt]{1-6}
			${\textit{Dataset 5}}$  & 4  &	2		& \parbox{1.8cm}{$n_0 = 948\\ n_1 = 948$}	& $\bm{\Sigma}_{\text{pooled}}$ & \parbox{4cm}{balanced sample sizes, homogeneous covariance matrices,
				same temporal trends of group means}  \\ 
			\arrayrulecolor{black}\bottomrule
		\end{tabular*}
		\label{comp_hours}
	\end{table}

	
	\vfill
	\begin{figure}[H]				
		\begin{subfigure}{\textwidth}
			\hspace{0.05cm}
			\includegraphics[scale = 0.22,trim={0cm, 0, 0cm, 0},clip]{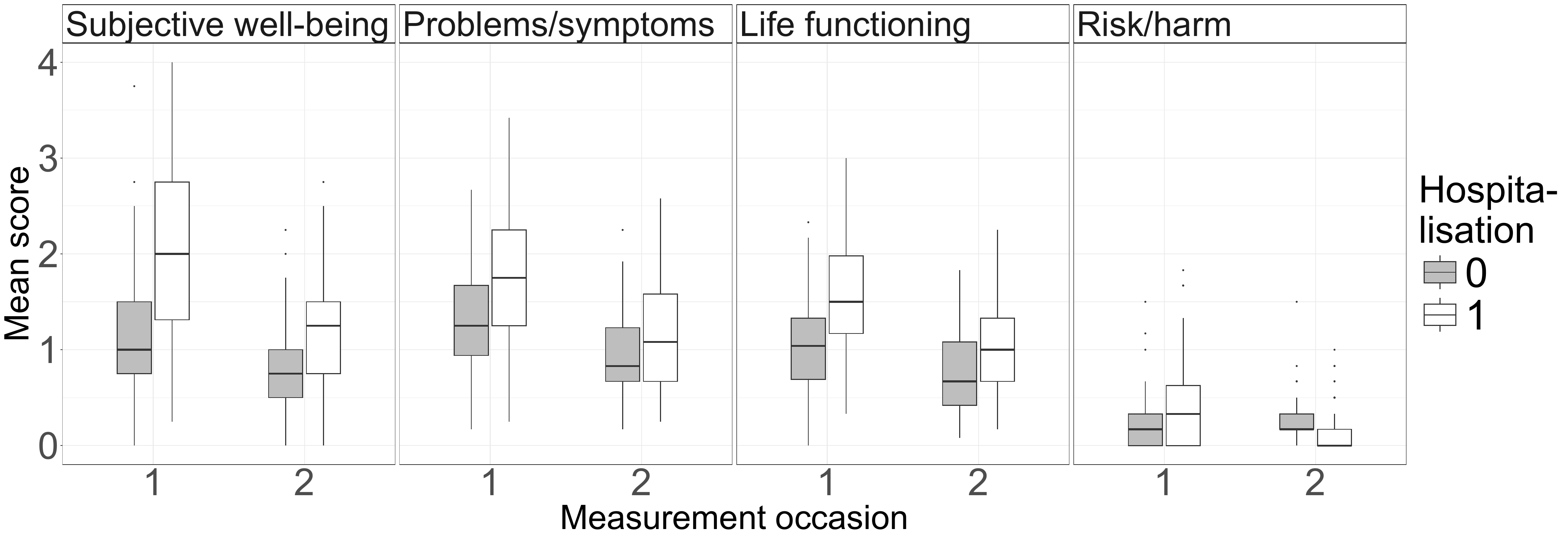}   
			\vspace{-0.57cm}
			\caption[]{}
			\vspace{-0.1cm}
		\end{subfigure}		
		\begin{subfigure}{\textwidth}
			\hspace{-0.05cm}		
			\includegraphics[scale = 0.22]{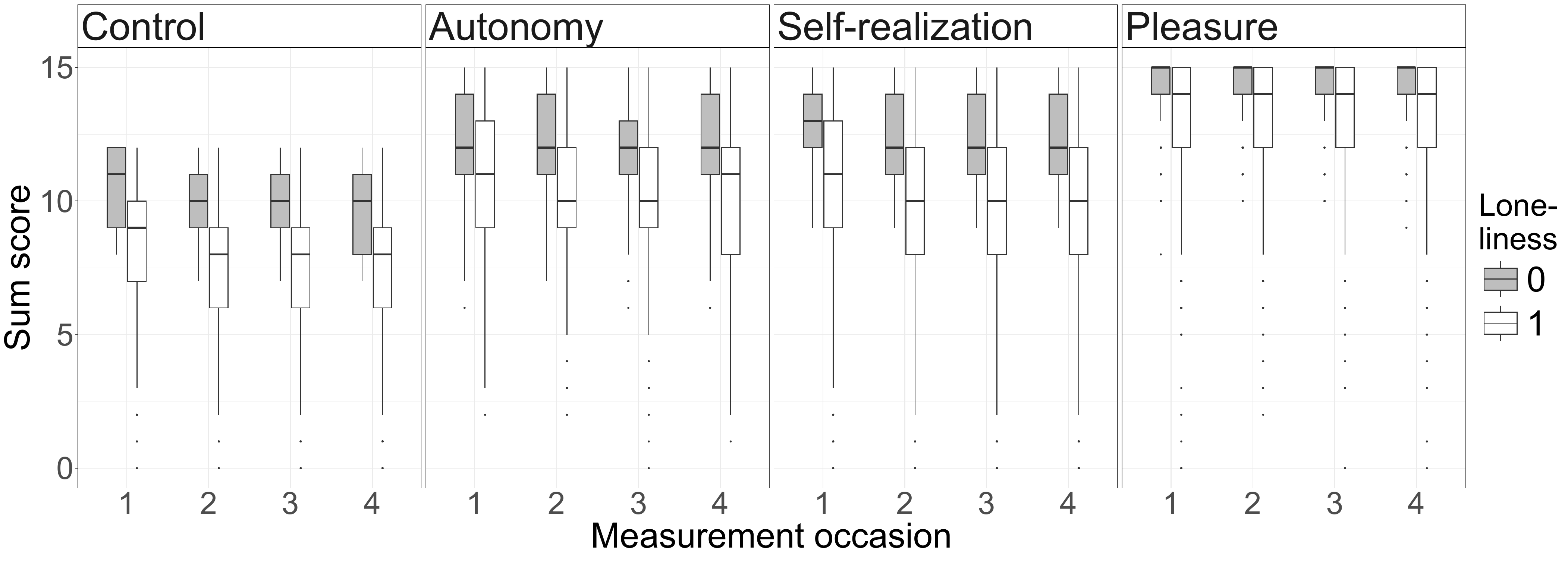} 
			\vspace{-0.57cm}
			\caption[]{}
		\end{subfigure}	
		\caption*{}
	\end{figure}
	
	\section{Methods}\label{methods}
	
	In the following section, we will describe the traditional repeated measures LDA, which relies on the multivariate normality assumption, its robust versions and the nonparametric longitudinal SVM for classification of nonnormally distributed repeated measures data. We will compare the performance of these methods in a neutral comparison study with respect to multiple performance measures. An overview of the considered methods is given in Table \ref{over}. Each classification method is considered in combination with or without previous outlier removal by trimming algorithms. An overview of the steps in the simulation study is shown in Table \ref{simproc}. Further details are included in Section \ref{sim_soft}.\\
	We consider a situation with a categorical outcome variable $y \in \{0,1\}$, where measurements of $d$ variables are taken at $t$ consecutive time points instead of only a single time point in $n = n_0 + n_1$ individuals. We consider complete data, i.e. for each individual $j \in \{1,\dots,n_i\}$, each measurement $l = 1,\dots,d$ is taken at each time point $k = 1,\dots,t$. The aim is to estimate a classification rule from the (training) data that can classify new observations (from separate independent test data) into one of two groups.\\ 
	
	\begin{table}[htb]
		\centering
		\small		
		\caption[Overview of the considered linear classification methods for nonnormally distributed multivariate repeated measures data.]{\small{Overview of the considered linear classification methods for nonnormally distributed multivariate repeated measures data. The performance of each classification method is estimated either without or in combination with preceding  
				multivariate outlier removal (using the Minimum Volume Ellipsoid (MVE) or the Minimum Covariance Determinant (MCD) algorithm, respectively).}}	
		\label{over}
		\begin{tabular*}{\textwidth}{@{}p{13em}p{22em}p{7em}@{}} \arrayrulecolor{black}\cmidrule[0.1pt]{1-3}
			Linear classification method & Description &  Abbreviation \\ \arrayrulecolor{black}\cmidrule[0.01pt]{1-3}		
			\parbox{13em}{Repeated measures linear discriminant analysis (LDA)\\ { (Section \ref{secLDA})}  \\  \\
				1) \hspace{0.1cm} standard/traditional \\ \\  \\
				2) \hspace{0.1cm} robust\\  \\  \\  \\  \\} & 
			\parbox{22em}{Parametric method depending on estimates of the group means and common covariance matrix \\ \\ \\ \hspace{0.05cm} (unstructured) pooled covariance matrix, \\ requires multivariate normality \\ \citep{Lix2010}   
				\\ 
				a) \hspace{0.05cm} (parsimonious) Kronecker product covariance estimated by flip-flop algorithm \citep{Brobbey2021}\\ 
				b)  (unstructured) covariance matrix estimated using the joint Generalized Estimating Equations model \citep{Brobbey2022} \\} 
			&   \parbox{7em}{\vspace{0.5cm}   LDA($\Sigma_{\text{pooled}}$)  \\ \\ \\ LDA($\Sigma_{\text{KP}}$)  \\ \\  LDA(GEE) }  \\ \arrayrulecolor{gray}\cmidrule[0.01pt]{1-3}\\[-1em]
			\parbox{13em}{Longitudinal Support Vector Machine (SVM)
				using a linear kernel \\  {(Section \ref{secSVM})} }  &
			\parbox{22em}{Nonparametric method independent of distributional assumptions \citep{Chen2011}\\  \\}		& \parbox{7em}{ SVM \\ \\ \\} 
			\\ \arrayrulecolor{black}\cmidrule[0.1pt]{1-3}
		\end{tabular*}
		
	\end{table}

	\subsection{Multivariate repeated measures LDA}\label{secLDA}
	For LDA, the unknown parameters $\bm{\mu}_i \in \mathds{R}^{dt}$, i.e. the group-specific mean vectors, and $\bm{\Sigma} \in \mathds{R}^{dt \times dt}$, i.e. the pooled covariance matrix, need to be estimated from the data  $\mathbf{X} = \{\mathbf{X}_{ij1}^T, \dots, \mathbf{X}_{ijt}^T\}_{\substack{i \in \{0,1\}\\j = 1,\dots,n_i}} \in  \mathds{R}^{n \times dt}$, where $\mathbf{X}_{ijk} \in \mathds{R}^{d}$ are continuous measurements. Here, $i \in \{0,1\}$ represents the group label, $j \in \{1,\dots, n_i\}$ the patient, $k \in \{1,\dots,t\}$ the time point, and $d$ the number of variables. The total sample size is denoted by $n = n_0 + n_1$. The covariance matrix $\bm{\Sigma} \in \mathds{R}^{dt \times dt}$ is assumed to be positive definite. The traditional LDA assumes multivariate normality of the data, $\mathbf{X}_{i} \stackrel{\text{iid}}{\sim} \mathcal{N}_{dt}(\bm{\mu}_i, \bm{\Sigma})$, as well as equality of group covariance matrices (homoscedasticity), $\bm{\Sigma}_0 = \bm{\Sigma}_1 = \bm{\Sigma}$.
	Brobbey et al. (\citeyear{Brobbey2021, Brobbey2022}) developed two approaches for robust LDA (when data deviate from multivariate normality) based on the Kronecker product estimate of the covariance matrix $\bm{\Sigma}$ that will be described in Section \ref{lda_trim} and Section \ref{lda_gee}. Here, we will briefly explain the rationale behind these modified LDA approaches and introduce the general LDA classification rule.\\
	Assuming that $\bm{\Sigma}$ is unstructured, all distinct correlations between each pair of the $d$ variables and each combination of the $t$ time points must be estimated. If the dataset is small, the estimate $\widehat{\bm{\Sigma}}$ may become singular, i.e. if $n \leq dt$. In order to reduce the complexity of $\bm{\Sigma}$ or to estimate $\bm{\Sigma}$ more efficiently, a reduced number of parameters can be considered by assuming, for example,  a Kronecker product structure $\bm{\Sigma} = \bm{\Sigma}_{t \times t}  \otimes \bm{\Sigma}_{d  \times d}$. Here, $\bm{\Sigma}_{t \times t} \in \mathds{R}^{t \times t}$ comprises the correlations between the $t$ time points and $\bm{\Sigma}_{d  \times d} \in \mathds{R}^{d \times d}$ comprises the correlations between the $d$ variables. The number of unknown parameters reduces from $(dt(dt+1)/2)$ for an unstructured covariance matrix to $d(d+1)/2 + t(t+1)/2$ for a Kronecker product covariance matrix \citep{Naik2001}.
	It can be estimated by the flip-flop algorithm, which gives maximum likelihood estimates of $\bm{\Sigma}_{t \times t}$ and $\bm{\Sigma}_{d  \times d}$  \citep{Lu2005}. The flip-flop algorithm is suitable in case each observation can be separated with respect to two factors, such as the time points and variables in case of multivariate longitudinal data.\\
	The LDA classification rule states that a new observation $\mathbf{X}_{ij} \in \mathds{R}^{dt}$ is assigned to class 0 if 
	\begin{equation*}
		\left(\mathbf{X}_{ij} - \frac{\bm{\mu}_0 + \bm{\mu}_1}{2}\right)^T \bm{\Sigma}^{-1} (\bm{\mu}_0 - \bm{\mu}_1) > \log \left(\frac{\pi_1}{\pi_0}\right) 
	\end{equation*}		
	where $\pi_i, i \in \{0,1\},$ is the prior probability of class $i$, $\bm{\mu}_0$ and $\bm{\mu}_1$ the respective group means, and  $\bm{\Sigma}^{-1}$ is the inverse covariance matrix \citep{Lix2010}. In the methods by Brobbey et al. (\citeyear{Brobbey2021, Brobbey2022}),  $\bm{\Sigma}^{-1}$ is replaced by $\bm{\Sigma}_{t \times t}^{-1} \otimes \bm{\Sigma}_{d \times d}^{-1}$.

	\subsubsection{Robust trimmed likelihood LDA for multivariate repeated measures data}\label{lda_trim}
	The rationale behind robust trimmed likelihood LDA for multivariate repeated measures data \citep{Brobbey2021} is to use more robust estimators of the sample mean and covariance matrix in order to increase the accuracy of LDA predictions in new data. Robust trimmed likelihood LDA for multivariate repeated measures data can also be used as a supporting analysis alongside the traditional LDA, showing that the results are not severely affected by outliers. \\
	Many estimators of these sample statistics are particularly prone to outliers, which are hard to detect in multivariate data with $d>2$ variables. 
	A popular measure of robustness, the finite sample breakdown point by Donoho (\citeyear{Donoho1982}) and Donoho and Huber (\citeyear{Donoho1983}), is the smallest number or fraction of extremely small or large values that must be added to the original sample that will result in an arbitrarily large value of the statistic. While many estimators of multivariate location and scatter break down when adding $n/(d+1)$ outliers \citep{Donoho1982}, estimators based on the Minimum Volume Ellipsoid (MVE) and Minimum Covariance Determinant (MCD) algorithms \citep{Rousseeuw1985} have a substantially higher break-down point of $(\lfloor n/2 \rfloor - d+1)/n$ \citep{Woodruff1993, Rousseeuw1999}.
	The high-breakdown linear discriminant analysis \citep{Hawkins1997} for cross-sectional data, for example, is  based on the MCD algorithm and has already been implemented in the \texttt{R} package \texttt{rrcov} \citep{Todorov2022}.\\
	The MCD is statistically more efficient than the MVE algorithm because it is asymptotically normal \citep{Butler1993}, its distances are more precise, i.e. it is more capable of detecting outliers \citep{Rousseeuw1999}.
	The MCD algorithm takes subsets of size $(n+d+1)/2 \leq h\leq n$ of the dataset (for $h>p$) and determines the particular subset of $h$ observations out of the $ \binom{n}{h} $ possible subsets for which the determinant of the sample covariance $\widehat{\bm{\Sigma}}$ becomes minimal. The MVE algorithm chooses the subset of $h$ observations for which the ellipsoid containing all $h$ data points becomes minimal.\\
	Brobbey (\citeyear{Brobbey2021}) suggests to estimate the class means $\bm{\mu}_0$ and $\bm{\mu}_1$ as well as the common covariance matrix $\bm{\Sigma}$ in the reduced dataset derived after applying the MCD or MVE algorithm, respectively. She furthermore suggests to  estimate the Kronecker product structure of the covariance matrix since it is more parsimonious than the unstructured equivalent, which may not be estimable for small sample sizes. We apply both versions, where we once estimate the unstructured pooled covariance matrix 
	\begin{equation*}
		\bm{\widehat{\Sigma}} = \frac{(n_0-1) \bm{\widehat{\Sigma}}_0 + (n_1-1) \bm{\widehat{\Sigma}}_1}{(n_0-1) + (n_1-1)}
	\end{equation*}
	and once the Kronecker product covariance $\bm{\widehat{\Sigma}} = \bm{\widehat{\Sigma}}_{t \times t}  \otimes \bm{\widehat{\Sigma}}_{d  \times d}$, where $\bm{\widehat{\Sigma}}_{t \times t}$ and $\bm{\widehat{\Sigma}}_{d  \times d}$ are the pooled covariances between the $t$ time points and $d$ variables, respectively. The flip-flop algorithm \citep{Lu2005} is used to estimate $\bm{\widehat{\Sigma}}_{t \times t}^{i}$ and $\bm{\widehat{\Sigma}}_{d  \times d}^{i}, i \in \{0,1\}$ from the data.\\

	\subsubsection{Generalized estimation equations (GEE) discriminant analysis for repeated measures data}\label{lda_gee} 
	\noindent Joint generalized estimating equations (GEEs) are another possibility to derive more robust estimates of the sample means and covariance matrix  from multivariate longitudinal data \citep{Brobbey2022, Inan2015}.
	GEEs provide population-level parameter estimates, which are consistent and asymptotically normally distributed even in case of misspecified working correlation structures of the outcome variables. The covariance matrix is estimated by a robust sandwich estimator \citep{Hardin2013}. 
	Brobbey et al. (\citeyear{Brobbey2022}) proposed the use of GEEs for multivariate repeated measures data in the context of repeated measures LDA as implemented by Inan (\citeyear{Inan2015}). The population-level estimates { ($\hat{\bm{\mu}}_0, \hat{\bm{\mu}}_1, \widehat{\bm{\Sigma}}$)} of the GEE model are plugged into the repeated measures LDA classification rule. For parsimony, the joint GEE model by Inan (\citeyear{Inan2015}) uses a decomposition of the working correlation matrix into a $t \times t$ within- and a $d \times d$ between-multivariate response correlation matrix through the Kronecker product.\\
	\noindent We fitted the joint GEE model by Inan (\citeyear{Inan2015}) to the data of each group $i \in \{0,1\}$ to obtain the class-specific means and covariance matrix estimates, which we subsequently pooled to obtain the common covariance matrix of the entire dataset. 
	Further details on the approach are given in Supplementary Material S.1.
	
	\subsection{Longitudinal Support Vector Machine}\label{secSVM}
	
	The original linear SVM for cross-sectional data and linearly separable classes  \citep{Vapnik1982} has been modified such that an overlap between the samples of both classes is to some extent allowed \citep{Cortes1995}. 
	\noindent Chen and Bowman (\citeyear{Chen2011}) further generalized this SVM classifier 
	such that it becomes applicable to longitudinal data. In their longitudinal SVM algorithm, temporal changes are modeled by considering a linear combination of the observations $\mathbf{X}_{ij} = \mathbf{x}_j \in \mathds{R}^{dt}$ and a parameter vector $\bm{\beta} = (1, \beta_1, \dots, \beta_{t-1}) $, which represents the coefficients for each time point $k$. Then, $ \widetilde{\textbf{x}}_{j} = \textbf{x}_{j1} + \beta_1 \textbf{x}_{j2} + \dots + \beta_{t-1} \textbf{x}_{jt}$, are provided as input to the traditional SVM. Combining the $d$ observations from all $t$ time points in a single vector assumes that the distances between time points are the same.
	The approach also assumes a fixed number of $d$ observations per time point $k$ (complete data) just as in case of LDA.
	Although this SVM classifier can also estimate nonlinear decision boundaries depending on the type of kernel matrix that is used, we apply a linear kernel in order to compare its performance to the other linear classifiers and since the absolute values of the weight vector 
	can be interpreted as variable importance in case of a linear kernel matrix.\\
	 	Although the SVM algorithm does not make any distributional assumptions, the regularization parameter $C$ needs to be optimized. We use the SSVMP algorithm \citep{Sentelle2016}, a modification of the SVMpath algorithm \citep{Hastie2004} to find the optimal value of $C$. The SSVMP algorithm is applicable for unequal class sizes and semidefinite kernel matrices in contrast to the original version by Hastie et al. (\citeyear{Hastie2004}). The path algorithm finds the optimal value $\lambda^{\scaleto{SVM}{3pt}} = 1/C$ with high accuracy, since it considers all possible values of $C$. At the same time, it is computationally efficient compared to the generally recommended grid search. It has been shown that the choice of $C$ can be critical for the generalizability of the SVM model \citep{Hastie2004}. \\
		The SSVMP algorithm \citep{Sentelle2015, Sentelle2016} optimizes the inverse of the regularization parameter, $\lambda^{\scaleto{SVM}{3pt}} = 1/C$. 	
		Starting with a high value of $\lambda^{\scaleto{SVM}{3pt}}$ such that all samples lie within the margin of the SVM, it successively determines a strictly decreasing sequence of  $\lambda^{\scaleto{SVM}{3pt}}$ values for which the set of support vectors changes for each $\lambda^{\scaleto{SVM}{3pt}}$ value, and it stops if no more observations are left inside of the margin (linearly separable case) or if the next $\lambda^{\scaleto{SVM}{3pt}}$  value would be zero.\\
		The longitudinal SVM algorithm by Chen and Bowman (\citeyear{Chen2011}) requires to specify a maximum number of iterations used for finding the optimal separating hyperplane parameters. In our case, the iterative algorithm for optimization of the Lagrange multipliers $\bm{\alpha}$ and temporal change parameters $\bm{\beta}$ in the longitudinal SVM is repeated until the Euclidean distance between two consecutive estimates of $\bm{\alpha}_m$ becomes less than $1\mathrm{E}$\textminus$08$ or the maximum number  of 100 iterative steps is reached.
		A summary of the longitudinal SVM algorithm using the linear soft-margin approach can be found in Supplementary Material S.2.    	
	
	\subsubsection{Nonparametric bootstrap approach}\label{method_boot}
		\noindent	The nonparametric bootstrap approach for point estimates by Wahl et al. (\citeyear{Wahl2016}) is an extension of the algorithm by Jiang et al. (\citeyear{Jiang2008}) and based on the .632+ bootstrap method \citep{Efron1997}, and thus assumes independence of  observations. It estimates the $.632+$ bootstrap estimate ($\hat{\theta}^{.632+}$) of the respective performance measure including a 95\% confidence interval.\\
		The .632+ bootstrap estimate   is computed as a weighted average of the apparent performance $\hat{\theta}^{orig,orig}$ (training and test data given by the original dataset) and the average ``out-of-bag'' (OOB) performance $\hat{\theta}^{bootstrap,OOB} = \sum\limits_{b=1}^B \hat{\theta}^{bootstrap,OOB}_b$ computed from $B$ bootstrap datasets (training data given by the bootstrap dataset, and test data given by the samples not present in the bootstrap dataset). The formula is:
		\begin{equation*}
			\hat{\theta}^{.632+} = (1-w) \cdot \hat{\theta}^{orig,orig} + w \cdot \hat{\theta}^{bootstrap,OOB},
		\end{equation*}
		where $w = \frac{0.632}{1-0.368 \cdot \text{r}}$ and r = $\frac{\hat{\theta}^{bootstrap,OOB} - \hat{\theta}^{orig,orig}}{\theta^{noinfo}   - \hat{\theta}^{orig,orig}}$. The value of $\theta^{noinfo}$ is 0.5 for predictive accuracy, sensitivity, and specificity. For the Youden index, this value is 0.\\ 
		Then each bootstrap dataset is assigned a weight $w_b =  \hat{\theta}^{bootstrap,bootstrap}_b - \hat{\theta}^{orig,orig}$, where $\hat{\theta}^{bootstrap,bootstrap}_b$ is the value of the performance measure, when the bootstrap dataset $b \in \{1,\cdots,B\}$ is used as training as well as test dataset. The  $\frac{\alpha^*}{2}$ and $1 - \frac{\alpha^*}{2}$ percentiles of the empirical distribution of these weights, $\xi_{\frac{\alpha^*}{2}}$ and $\xi_{1 - \frac{\alpha^*}{2}}$ , give the confidence interval of $\hat{\theta}^{.632+}$:
		\begin{equation*}
			[\hat{\theta}^{.632+} -  \xi_{1 - \frac{\alpha^*}{2}}, \hat{\theta}^{.632+} + \xi_{\frac{\alpha^*}{2}}]
	\end{equation*}	
	\subsection{Performance measures}\label{perf_meas}
		In order to compare class prediction of the classification algorithms in the independent test data, we used predictive accuracy, the Youden index, sensitivity, and specificity as measures of discrimination.
		Predictive accuracy is the number of correctly classified samples divided by the total number of samples. Sensitivity, or true positive rate, is the proportion of individuals among all individuals that have been predicted to belong to class 1, whose class prediction matches their true class label. Specificity, or true negative rate, is the the proportion of individuals among all individuals that have been predicted to belong to class 0, whose class prediction matches their true class label. The  Youden index \citep{Youden1950} combines  sensitivity and specificity of the classification model into a single measure (Youden index = $|$Sensitivity + Specificity - 1$|$). \\
		Recommendations based on theses measures can differ a lot. Predictive accuracy of an algorithm may be high in data with highly unbalanced classes if the label of the larger class is predicted for all samples. In this case the Youden index will have the minimum value of zero. Therefore it is reasonable to consider both measures, predictive accuracy and the Youden index. 
	
	\subsection{Simulation study approach and software}\label{sim_soft}
	
	Our simulation study aims at mimicking  reference datasets from psychological applications. See Section \ref{ref_data} for a detailed description of these datasets.
	A brief overview of the steps in the simulation study is given in Figure \ref{simproc}. For each scenario, 2000 datasets are simulated. Sample sizes for the training data are chosen identical to the sample sizes of the reference datasets. Sample sizes for the test data for each group are 10 times the number of the  respective original group sample size in order to maintain the group size ratio.  A larger test sample size can be chosen in simulations since they do not rely on actual data. Variance in the performance estimates may thereby be decreased.   Data are simulated from the multivariate normal distribution (as a reference), from the multivariate truncated normal distribution which only takes on values within specified boundaries similar to the sum or mean scores in the reference data, respectively, and from the multivariate lognormally distributed data in order to include an extremely skewed distribution (overview in Table \ref{distpar}). Parameters needed for data simulations are estimated from the reference datasets (i.e. the pooled covariance matrix $\bm{\Sigma}$, or the group covariance matrices  $\bm{\Sigma}_0$ and $\bm{\Sigma}_1$, respectively, group means $\bm{\mu}_0$, and $\bm{\mu}_1$, and the lower and upper boundaries, \textbf{a} and \textbf{b}, of the sum or mean score, respectively).  Training data are either not trimmed or trimmed using the MCD and the MVE algorithm, respectively, keeping 90\% of the samples, before applying the classification algorithms.  In contrast to Brobbey at al. (\citeyear{Brobbey2021, Brobbey2022}), we did not use the restrictive assumption of a Kronecker product covariance structure for simulating the data. In contrast to Chen and Bowman (\citeyear{Chen2011}), the datasets to which we applied the method are not balanced in sample size. We would like to examine the methods' performance in more general simulation settings. \\ 
		 Since the SVM algorithm relies on the Euclidean distance to determine the optimal decision boundary, standardization is required as a data-preprocessing step. We standardized the data variable-wise (across time points) before applying the method. Centering and scaling is done using the \texttt{preProcess} function in the \texttt{R} package \texttt{caret} \citep{Kuhn2024}. More specifically, each training dataset is centered, and scaled to unit variance, and the same parameters are then used to standardize the test dataset in the same way \citep{Hsu2003}. 
	Machine-learning algorithms generally require the optimization of hyperparameters. 	Application of the linear SVM algorithm requires finding the optimal value of the hyperparameter $C$ which determines the maximum amount of overlap allowed between samples of both classes. We applied the simple SVM path (SSVMP) algorithm by Sentelle et al. (\citeyear{Sentelle2016}) as suggested by Chen and Bowman (\citeyear{Chen2011}) in order to determine the optimal regularization parameter $C$. It is available as \texttt{MATLAB} code \citep{Sentelle2015}, which we rewrote in \texttt{R}. Computation of the longitudinal SVM results including the computation of the optimal $C$ could only be done for the two smaller datasets (dataset 1 and dataset 3) due to limitations by computational complexity. \\
	The flip-flop algorithm  \citep{Lu2005} used by Brobbey (\citeyear{Brobbey2021}) for estimating the Kronecker product structure of the covariance matrix from the training data (for the LDA($\bm{\Sigma}_{\text{KP}}$) algorithm) was iterated until the Frobenius norm of two consecutive Kronecker product covariance matrices became less than or equal to $1\mathrm{E}$\textminus$04$, a proposed stopping criterion by Castaneda and Nossek (\citeyear{Castaneda2014}).\\
	We used the following software for data simulations. We implemented the longitudinal SVM in \texttt{R} using the \texttt{R} package \texttt{Rcplex} \citep{Bravo2021}.
	We used the implementations of the MVE and MCD algorithm from the \texttt{R} package \texttt{MASS} \citep{Ripley2022}, the joint GEE model as implemented in the \texttt{R} package \texttt{JGEE} \citep{Inan2015}, and implemented the version of the flip-flop algorithm in \texttt{R} as described in Lu and Zimmerman (\citeyear{Lu2005}).
	For simulation of multivariate normally, lognormally, and truncated normally distributed data, we used the respective functions from the \texttt{R} packages  \texttt{MASS} \citep{Ripley2022},  \texttt{compositions} \citep{Boogaart2022}, and  \texttt{tmvtnorm} \citep{Wilhelm2022}.  For the truncated normal distribution, the  rejection method (default) was used.\\
	
	\begin{table}[htb]
		\centering
		\small		
		\caption[Parameterizations of the multivariate distributions used for data simulations.]{\small{Parameterizations of the multivariate distributions for group $i \in \{0,1\}$. The multivariate truncated normal distribution is defined by lower and upper boundaries, $\textbf{a} \in \mathds{R}^{dt}$ and  $\textbf{b}  \in \mathds{R}^{dt}$, respectively, in addition to the mean ($\bm{\mu}_{i}$) and covariance ($\bm{\Sigma}$) parameters.}}
		\label{distpar}				
		\begin{tabular*}{0.7\textwidth}{@{}p{16em}p{13.7em}@{}} \toprule 
			Distribution	& {Parameterization}  \\ \midrule		
			Multivariate normal  											& $\mathcal{N}_{dt}(\bm{\mu}_{i}, \bm{\Sigma})$	\\ \arrayrulecolor{gray!50}\cmidrule[0.01pt]{1-2}
			\\[-1em]
			Multivariate lognormal 											& $\mathcal{LN}_{dt}(\bm{\mu}_{i}, \bm{\Sigma})$    \\ \arrayrulecolor{gray!50}\cmidrule[0.01pt]{1-2}
			\\[-1em]
			Multivariate truncated normal 									& $\mathcal{TN}_{dt}(\bm{\mu}_{i}, \bm{\Sigma}, \textbf{a}, \textbf{b})$  \\ \arrayrulecolor{black}\bottomrule
		\end{tabular*}
		
	\end{table}

	\definecolor{grey}{RGB}{217, 217, 217}
	\definecolor{lightgrey}{RGB}{230, 230, 230}
	\definecolor{darkgrey}{RGB}{100, 100, 100}
	\definecolor{black}{RGB}{51, 51, 51}
	\definecolor{white2}{RGB}{0, 0, 0}
	
	\begin{figure}[htb]
		\renewcommand\thefigure{3}	
		\begin{tikzpicture}[>=stealth,sloped,auto,  
			font=\sffamily,		
			N11/.style = {rectangle, rounded corners, minimum width=3.5cm, minimum height=0.7cm, text centered, font=\small, color=black, draw=grey, line width=1, fill=lightgrey,inner sep=6pt},
			N21/.style = {rectangle, rounded corners, minimum width=1.2cm, minimum height=0.7cm, text centered, font=\small, color=black, draw=white2, line width=0.1, fill=white},
			N24/.style = {rectangle, rounded corners, minimum width=2cm, minimum height=1cm, text centered, font=\small, color=black, draw=white2, line width=0.1, fill=white}]
			\tikzset{every node}=[font=\small\sffamily]
			
			\begin{scope}[]
				\matrix[nodes={rectangle, draw, align=left, anchor=center}, column sep=1.5cm,
				column 1/.style={anchor=base },
				column 2/.style={anchor=base},
				column 3/.style={anchor=base },
				column 4/.style={anchor=base },
				column 5/.style={anchor=base }
				]{
					\node [N11](Parameters) {Parameters\strut}; & 
					\node [N24](est) {estimate $\bm{\Sigma}$ (or $\bm{\Sigma}_0$ and $\bm{\Sigma}_1$), $\bm{\mu}_0, \bm{\mu}_1$ from reference data \\(mean or sum scores of Likert-type responses)\strut, \\ determine lower and upper boundary (\textbf{a}, \textbf{b}) of the score \strut}; &
					&
					&
					\\};
			\end{scope}
			
			\begin{scope}[yshift = -1.75cm]
				\matrix[nodes={rectangle, draw, align=left, anchor=center}, column sep=1.5cm,nodes in empty cells,
				column 1/.style={anchor=base },
				column 2/.style={anchor=base },
				column 3/.style={anchor=base },
				column 4/.style={anchor=base },
				column 5/.style={anchor=base }]{
					\hspace{-0.8cm}	 \node [N11](DataSim) {Data simulation\strut}; &   
					\hspace{-2.125cm} \node [N21](N) {$\mathcal{N}_{dt}(\bm{\mu}_i, \bm{\Sigma})$} ; & 
					\hspace{-4.525cm} 		\node [N21](LN) {$\mathcal{LN}_{dt}(\bm{\mu}_i, \bm{\Sigma})$}; &  
					\hspace{-9.575cm}		\node [N21](TN) {$\mathcal{TN}_{dt}(\bm{\mu}_i, \bm{\Sigma}, \textbf{a}, \textbf{b})$}; &\\};  
			\end{scope}
			
			\begin{scope}[yshift = -3.5cm]
				\matrix[nodes={rectangle, draw, align=left, anchor=center}, column sep=1.5cm,nodes in empty cells,
				column 1/.style={anchor=base },
				column 2/.style={anchor=base },
				column 3/.style={anchor=base },
				column 4/.style={anchor=base },
				column 5/.style={anchor=base }]{
					\hspace{-2.25cm}	 \node [N11](Trimming) {Trimming \\(of training data)\strut}; & 
					\hspace{-4.2cm} \node [N21](none) {none} ; & 
					\hspace{-8.5cm} 		\node [N21](MVE) {MVE}; &
					\hspace{-17.1cm}		\node [N21](MCD) {MCD}; &\\};
			\end{scope}
			
			\begin{scope}[yshift = -5.75cm]
				\matrix[nodes={rectangle, draw, align=left, anchor=center}, column sep=1.5cm,nodes in empty cells,
				column 1/.style={anchor=base },
				column 2/.style={anchor=base },
				column 3/.style={anchor=base },
				column 4/.style={anchor=base },
				column 5/.style={anchor=base }]{
					\hspace{-0.55cm} \node [N11] {Classification \\ (parameter estimation\\ in training data,\\ prediction in test data)\; \strut}; & 
					\hspace{-1.55cm} \node [N21](pool) {LDA($\bm{\Sigma}_{\text{pooled}}$)} ; & 
					\hspace{-4.05cm} \node [N21](KP) {LDA($\bm{\Sigma}_{\text{KP}}$)}; &
					\hspace{-9.05cm} \node [N21](GEE) {LDA(GEE)}; &
					\hspace{-19.1cm} \node [N21](SVM) {SVM}; \\};
			\end{scope}
			
			\begin{scope}[yshift = -7.75cm]
				\matrix[nodes={rectangle, draw, align=left, anchor=center}, column sep=1.5cm,
				column 1/.style={anchor=base },
				column 2/.style={anchor=base},
				column 3/.style={anchor=base },
				column 4/.style={anchor=base },
				column 5/.style={anchor=base }]{
					\hspace{-0.95cm} \node [N11](PerfMeas) {Performance measures\strut}; & 
					\hspace{-0.92cm} \node [N24](text) {compute predictive accuracy, Youden index,\\ sensitivity, specificity}; &
					&
					&
					\\};
			\end{scope}

			\draw[->, draw = darkgrey] (-2.25,-0.69) -- (-3.25,-1.4);  
			\draw[->, draw = darkgrey] (-0.2,-0.69) -- (-0.2,-1.4);
			\draw[->, draw = darkgrey] (2.175,-0.69) -- (3.175,-1.4);
			
			\draw[->, draw = darkgrey] (-3.3,-2.1) -- (-2.85,-3.15);
			\draw[->, draw = darkgrey] (-3.3,-2.1) -- (-0.3,-3.15);
			\draw[->, draw = darkgrey] (-3.3,-2.1) -- (2.35,-3.15);
			\draw[->, draw = darkgrey] (-0.2,-2.1) -- (-2.85,-3.15);
			\draw[->, draw = darkgrey] (-0.2,-2.1) -- (-0.3,-3.15);
			\draw[->, draw = darkgrey] (-0.2,-2.1) -- (2.35,-3.15);
			\draw[->, draw = darkgrey] (3.175,-2.1) -- (-2.85,-3.15);
			\draw[->, draw = darkgrey] (3.175,-2.1) -- (-0.3,-3.15);
			\draw[->, draw = darkgrey] (3.175,-2.1) -- (2.35,-3.15);
			
			\draw[->, draw = darkgrey] (-2.85,-3.85) -- (-3.05,-5.4);
			\draw[->, draw = darkgrey] (-2.85,-3.85) -- (1.8,-5.4);
			\draw[->, draw = darkgrey] (-2.85,-3.85) -- (3.75,-5.4);
			\draw[->, draw = darkgrey] (-2.85,-3.85) -- (-0.505,-5.4);
			\draw[->, draw = darkgrey] (-0.3,-3.85) -- (-0.505,-5.4);
			\draw[->, draw = darkgrey] (-0.3,-3.85) -- (-3.05,-5.4);
			\draw[->, draw = darkgrey] (-0.3,-3.85) -- (1.8,-5.4);
			\draw[->, draw = darkgrey] (-0.3,-3.85) -- (3.75,-5.4);
			\draw[->, draw = darkgrey] (2.35,-3.85) -- (1.8,-5.4);
			\draw[->, draw = darkgrey] (2.35,-3.85) -- (3.75,-5.4);
			\draw[->, draw = darkgrey] (2.35,-3.85) -- (-0.505,-5.4);
			\draw[->, draw = darkgrey] (2.35,-3.85) -- (-3.05,-5.4);
			
			\draw[->, draw = darkgrey] (-3.05,-6.1) -- (-2.4,-7.25);
			\draw[->, draw = darkgrey] (-0.505,-6.1) -- (-0.6,-7.25);
			\draw[->, draw = darkgrey] (1.8,-6.1) -- (1.2,-7.25);
			\draw[->, draw = darkgrey] (3.75,-6.1) -- (3,-7.25);	 
		\end{tikzpicture}
		
		\caption[Overview of the steps in the simulation study.]{\fontsize{10}{13}\selectfont{Overview of the steps in the simulation study for a particular reference dataset.}\\
			{\fontsize{10}{13}\selectfont Abbreviations: $\mathcal{N}_{dt}$ - multivariate normal distribution  $\mathcal{LN}_{dt}$ - multivariate lognormal distribution, $\mathcal{TN}_{dt}$ - multivariate truncated normal distribution, $d$ - $\#$ variables, $t$ - $\#$ time points, LDA($\Sigma_{\text{pooled}}$) - Linear discriminant analysis (pooled covariance matrix), LDA($\Sigma_{\text{KP}}$) - Linear discriminant analysis  (Kronecker product covariance matrix), LDA(GEE) - Linear discriminant analysis (covariance matrix based on generalized estimating equations estimates), SVM - longitudinal Support vector machine, MVE - minimum volume ellipsoid algorithm, MCD - minimum covariance determinant algorithm.}}
		\label{simproc}
	\end{figure}

	\clearpage	
	\section{Results and discussion}\label{results}
	
	\subsection{Performance in the reference data}
	
\begin{figure}[H]	
		\renewcommand\thefigure{4}				
		\begin{subfigure}{\textwidth}
			\begin{minipage}{.45\textwidth}
				\includegraphics[scale = 0.5]{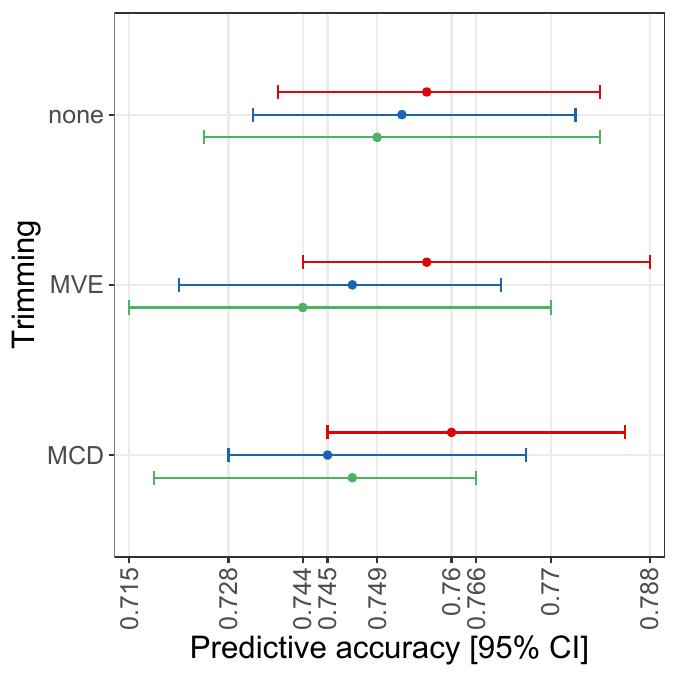}
			\end{minipage}
			\begin{minipage}{.03\textwidth}
				\hspace{0.05cm}
				\vspace{5.5cm}
				\caption*{(e)}
			\end{minipage}
			\begin{minipage}{.4\textwidth}
				\includegraphics[scale = 0.5]{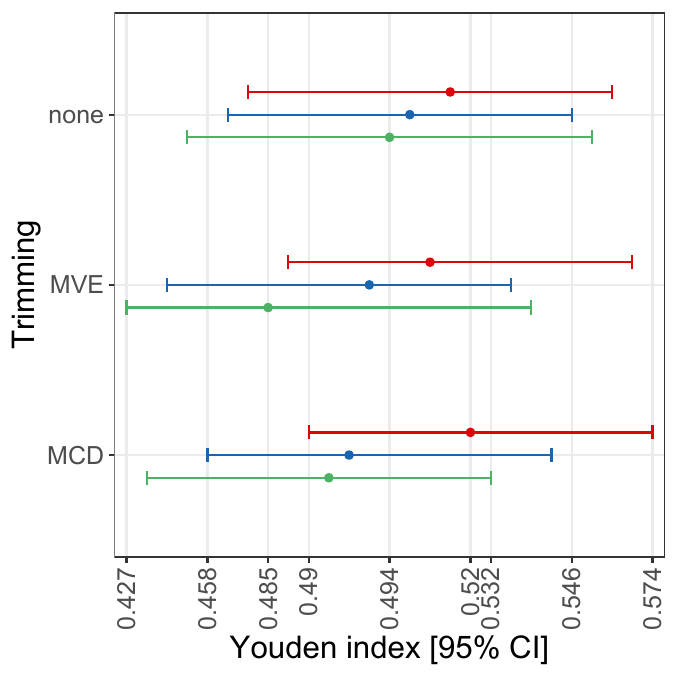}
			\end{minipage}
			\begin{minipage}{.01\textwidth}
				\includegraphics[scale = 0.6, trim={6cm 6cm 6cm 6cm},clip]{Fig4_boot_CI_4_legend.pdf}
			\end{minipage}
		\end{subfigure}	
		\caption[Performance in the reference data using the bootstrap approach by Wahl et al. (\citeyear{Wahl2016}) and 2000 bootstrap datasets: $\hat{\theta}^{.632+}$  and respective 95\% confidence intervals for the performance measures predictive accuracy and Youden index.]{\fontsize{9}{12}\selectfont
			{ Performance in the reference data using the bootstrap approach by Wahl et al. (\citeyear{Wahl2016}) and 2000 bootstrap datasets: $\hat{\theta}^{.632+}$  and respective 95\% confidence intervals for the performance measures predictive accuracy and Youden index. \\
					(a) Dataset 1: CORE-OM dataset, group variable \textit{hospitalisation} ($n_0 = 42, n_1 = 142$)\\
					(b) Dataset 2: CASP-19 dataset, group variable \textit{loneliness} ($n_0 = 948, n_1 = 1682$) \\
					(c) Dataset 3: modified Dataset 1, such that group means collapsed over time points are equal, and group means have opposite temporal trends\\
					(d) Dataset 4: modified Dataset 2 (time points 1 \& 2), such that group means are equal, and group covariance matrices differ,\\
					(e) Dataset 5: modified Dataset 2 (time points 1 \& 2), balanced class sizes by random undersampling of group 1.} 
		}
		\label{boot_ci}
	\end{figure}

\noindent For computing point estimates of the performance measures including confidence intervals in the reference data, we used the bootstrap approach described in Section \ref{method_boot}. 
		Estimates of predictive performance and the Youden index are shown in Figure \ref{boot_ci}, those of sensitivity and specificity can be found in Figure S 2. The bootstrap estimates and their respective confidence intervals are also shown in Table S 3.\\
		Figure \ref{boot_ci} shows that the two methods LDA($\bm{\Sigma}_{\text{pooled}}$) and LDA($\bm{\Sigma}_{\text{KP}}$) have a very similar performance in all scenarios, and generally perform best. Including Figure S 2, these two methods tend to have more moderate values of sensitivity and specificity even for highly imbalanced datasets (datasets 1,2,3). However, similar to LDA(GEE), they are (almost) incapable to accurately predict the correct class of individuals from the minority class when group means are identical and only group covariance matrices differ (dataset 4). All three methods, LDA($\bm{\Sigma}_{\text{pooled}}$), LDA($\bm{\Sigma}_{\text{KP}}$), and LDA(GEE),  predominantly predict that individuals belong to the majority class in this scenario, probably because its covariance matrix has a greater weight when computing the inverse of the pooled covariance matrix for the classification rule (Section \ref{secLDA}). In comparison, LDA(GEE) and SVM perform worse for unequal class sizes, of which SVM performs worse compared to LDA(GEE), particularly because its specificity (prediction of the minority class) is very low. Comparing the performance for dataset 1 (same temporal trends of group means) and dataset 3 (opposite temporal trends of group means), the results of all performance measures considerably improve for LDA($\bm{\Sigma}_{\text{pooled}}$) and LDA($\bm{\Sigma}_{\text{KP}}$).  For LDA(GEE) there is almost no change (very slight improvement), and overall no difference for the SVM. Results for dataset 5 show that using balanced instead of the imbalanced data (dataset 2), increase specificity of all LDA methods but particularly for LDA(GEE), resulting in a higher Youden index. Trimming of the training data does only in some cases improve the performance in the test data. A slight improvement of predictive performance and Youden index at the same time can only be observed in some cases: for LDA($\bm{\Sigma}_{\text{pooled}}$) when applied to dataset 1 (after MVE trimming), dataset 3 (after MVE and MCD trimming), dataset 5 (after MCD trimming), for LDA($\bm{\Sigma}_{\text{KP}}$) when applied to dataset 1 (MVE trimming), dataset 3 (MVE and MCD trimming), and LDA(GEE) when applied to dataset 1 (MCD trimming).

	\begin{figure}[H]				
		\begin{subfigure}{\textwidth}
			\begin{minipage}{.45\textwidth}
				\includegraphics[scale = 0.5]{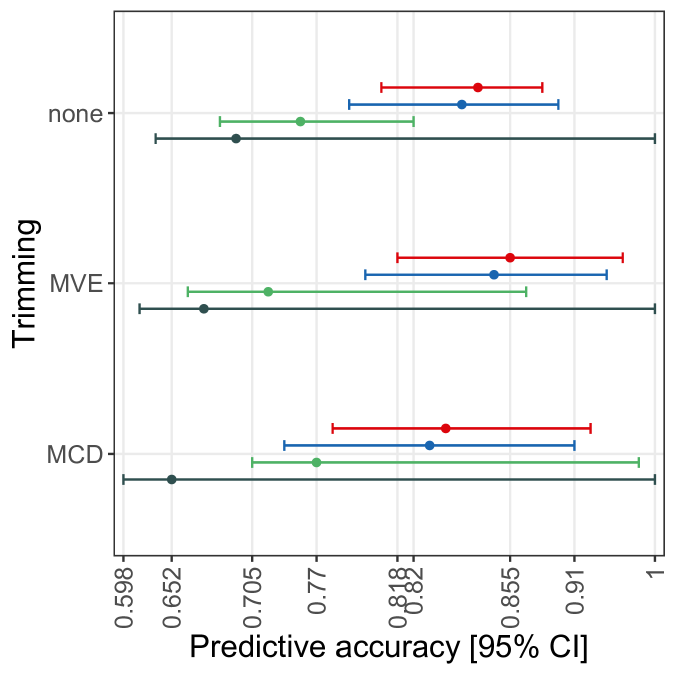}
			\end{minipage}
			\begin{minipage}{.03\textwidth}
				\hspace{0.05cm}
			\end{minipage}
			\begin{minipage}{.4\textwidth}
				\includegraphics[scale = 0.5]{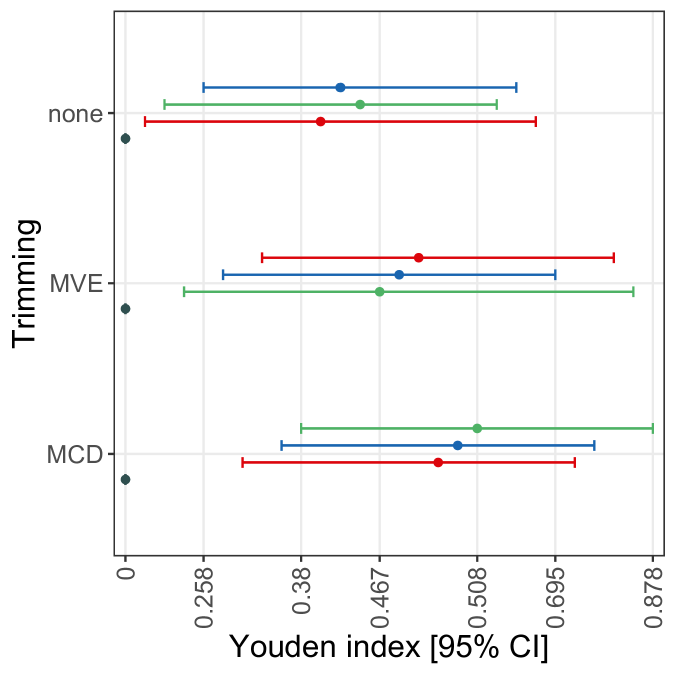}
			\end{minipage}
			\vspace{-0.55cm}
			\caption*{\hspace{-0.8cm}(a)}
		\end{subfigure}	
		\begin{subfigure}{\textwidth}
			\begin{minipage}{.45\textwidth}
				\vspace{-1.5cm}
				\includegraphics[scale = 0.5]{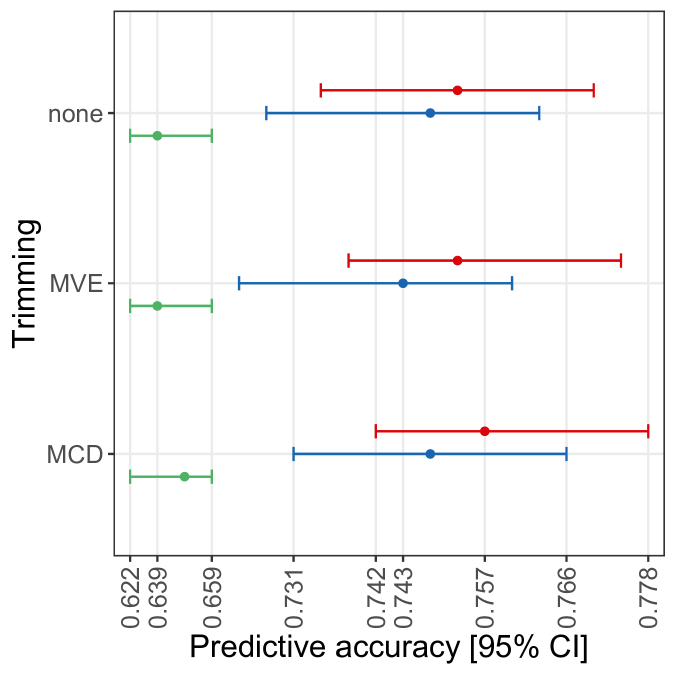}
			\end{minipage}
			\begin{minipage}{.03\textwidth}
				\vspace{3.75cm}
				\caption*{(b)}
			\end{minipage}
			\begin{minipage}{.4\textwidth}
				\vspace{-1.5cm}
				\includegraphics[scale = 0.5]{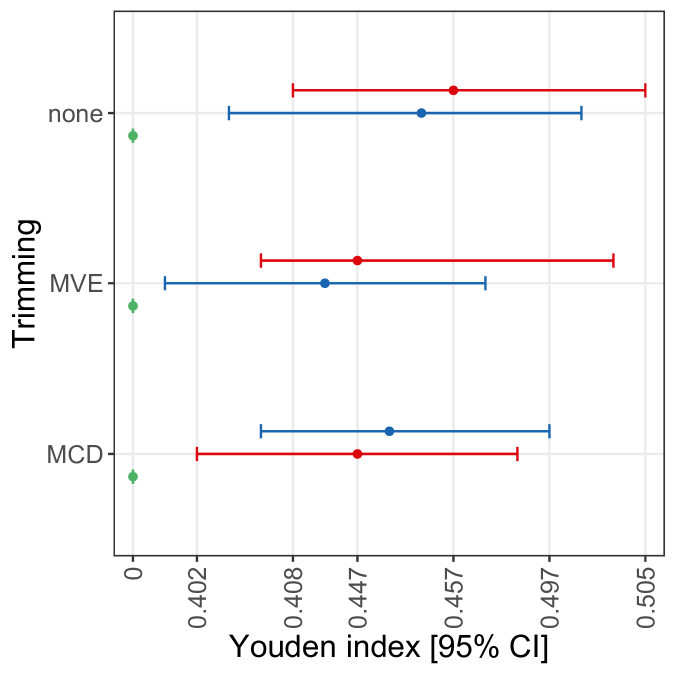}
			\end{minipage}
			\begin{minipage}{.01\textwidth}
				\includegraphics[scale = 0.6, trim={6cm 6cm 6cm 6cm},clip]{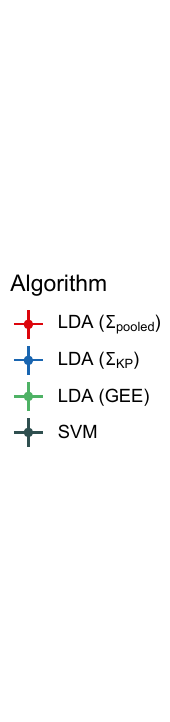}
			\end{minipage}				
		\end{subfigure}	
		\begin{subfigure}{\textwidth}
			\begin{minipage}{.45\textwidth}
				\vspace{-1.5cm}
				\includegraphics[scale = 0.5]{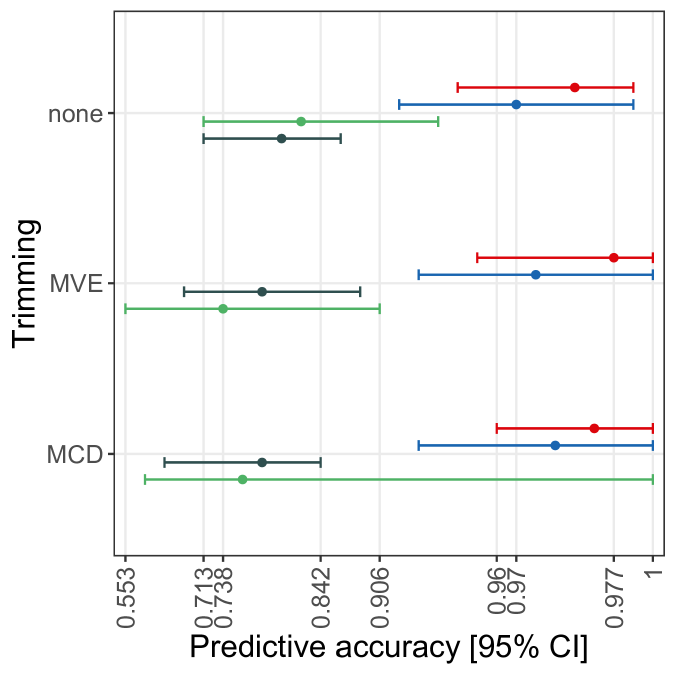}
			\end{minipage}
			\begin{minipage}{.03\textwidth}
				\hspace{0.05cm}
			\end{minipage}
			\begin{minipage}{.4\textwidth}
				\vspace{-1.5cm}
				\includegraphics[scale = 0.5]{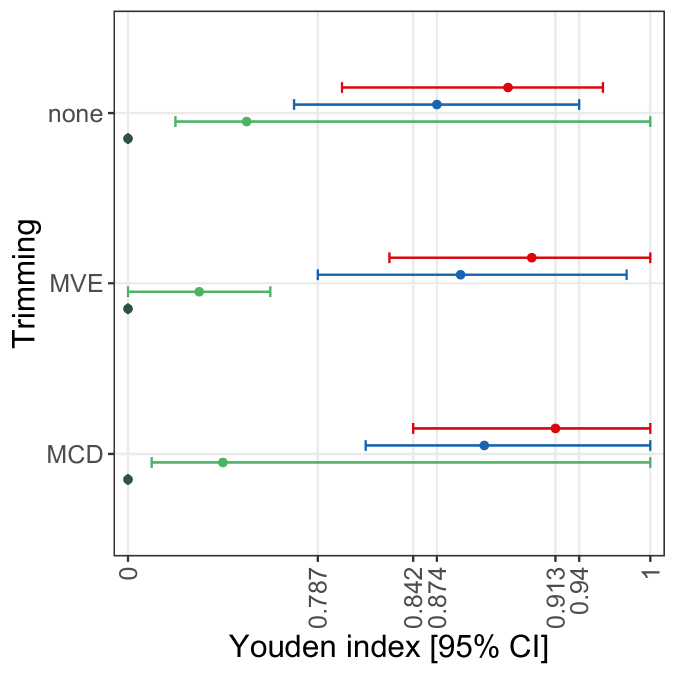}
			\end{minipage}
			\vspace{-0.55cm}
			\caption*{\hspace{-0.8cm}(c)}
		\end{subfigure}
		\begin{subfigure}{\textwidth}
			\begin{minipage}{.45\textwidth}
				\includegraphics[scale = 0.5]{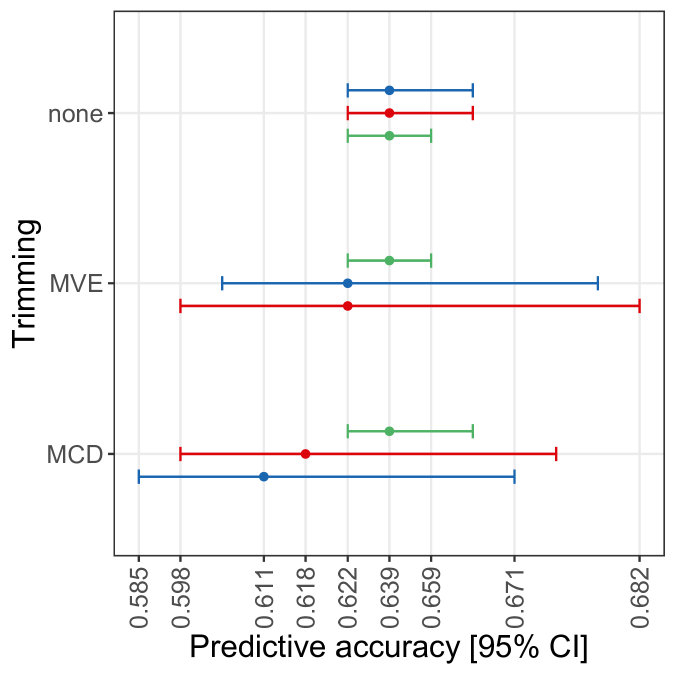}
			\end{minipage}
			\begin{minipage}{.03\textwidth}
				\hspace{0.05cm}
			\end{minipage}
			\begin{minipage}{.4\textwidth}
				\includegraphics[scale = 0.5]{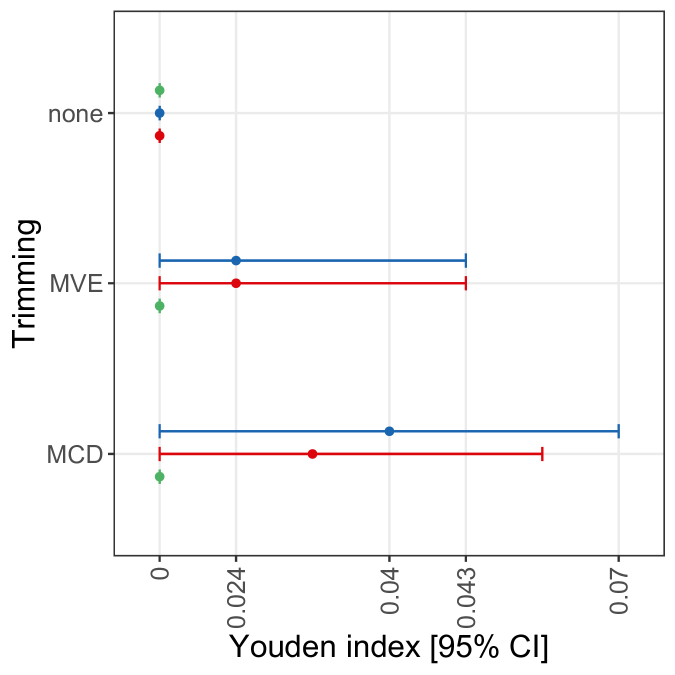}
			\end{minipage}
			\vspace{-0.55cm}
			\caption*{\hspace{-0.8cm}(d)}
		\end{subfigure}
		\caption*{}
	\end{figure}
	
	\subsection{Performance in the simulated data}
	
	{	For data simulations we assumed homogeneity of covariance matrices (which is a LDA assumption) for data generation based on datasets 1, 2, 3, and 5, despite heterogeneity in the reference datasets. Figure S 1 shows plots for comparison of the components of Box's M test, which is known to be very sensitive to violations of the normality assumption, and results may therefore not be reliable. Log determinants and log eigenvalues of the covariance matrices differ from each other suggesting heterogeneity of covariances in the reference data. Only for dataset 4 we assumed heterogeneous covariance matrices for data generation in order to compare the methods' performance under violation of this assumption when group means are identical at the same time.\\
		The second LDA assumption is multivariate normality of the data. Table S 4 shows that lognormally distributed multivariate data differ most from multivariate normality according to the Mardia measure of multivariate skewness (highest absolute number of significant test results). Truncated normally distributed data differ more significantly from multivariate normality for larger sample sizes and/or a higher number of measurement occasions. Especially for datasets 2 and 4, respectively, trimming the data  using the MCD algorithm notably decreases deviation from multivariate normality in truncated normally distributed data, which is also true for datasets 1 and 3, respectively, when the MCD algorithm is applied to the lognormally distributed data. This effect is weaker for the MVE algorithm. This shows at least that the MCD algorithm, which has been found to be more suitable for outlier detection compared to the MVE algorithm \citep{Rousseeuw1999}, may be useful in case outliers or non-normality is assumed to bias parameter estimates. On the other hand, the optimal trimming value has to be chosen in order  to not remove valuable observations from the data. There currently are no general guidelines. \\
		Table \ref{comp_hours} shows the computational times per algorithm, averaged over scenarios using different data distributions and trimming approaches. The method LDA($\Sigma_{\text{pooled}}$) has the advantage of

		\begin{table}[htb]
			\centering
			\small	
			\caption[Averaged computational times (hours) per algorithm.]{\small{Computational times (hours) per algorithm averaged over the simulated datasets  per reference dataset (irrespective of the data distribution and irrespective whether trimming has been done before application of the classification algorithm).\\ 		
					Dataset 1: CORE-OM dataset, group variable \textit{hospitalisation} ($n_0 = 42, n_1 = 142$)\\
					Dataset 2: CASP-19 dataset, group variable \textit{loneliness} ($n_0 = 948, n_1 = 1682$) \\
					Dataset 3: modified Dataset 1, such that group means collapsed over time points are equal, and group means have opposite temporal trends\\
					Dataset 4: modified Dataset 2 (time points 1 \& 2), such that group means are equal, and group covariance matrices differ,\\
					Dataset 5: modified Dataset 2 (time points 1 \& 2), balanced class sizes by random undersampling of group 1.}\\
				{  \footnotesize  Abbreviations: LDA($\Sigma_{\text{pooled}}$) - Linear discriminant analysis (pooled covariance matrix), LDA($\Sigma_{\text{KP}}$) - Linear discriminant analysis  (Kronecker product covariance matrix), LDA(GEE) - Linear discriminant analysis (covariance matrix based on generalized estimating equations estimates), SVM - Support vector machine.}} 
			\begin{tabular*}{\textwidth}{@{}p{7em}p{8em}p{8em}p{8em}p{8em}@{}} \toprule		
				&  \parbox{1.2cm}{ LDA($\bm{\Sigma}_{\text{pooled}}$)} &  \parbox{1.2cm}{ LDA($\bm{\Sigma}_{\text{KP}}$)}    & \parbox{1.2cm}{LDA(GEE)}  		&  SVM   \\ \midrule
				${\textit{Dataset 1}}$  & 0.08  &	1.05		& 0.34	& 64.29    \\  \arrayrulecolor{gray!50}\cmidrule[0.05pt]{1-5}
				${\textit{Dataset 2}}$  & 1.4  & 29.62         & 26.71 & \hspace{0.1cm} \textminus  \\ \arrayrulecolor{gray!50}\cmidrule[0.05pt]{1-5}
				${\textit{Dataset 3}}$  & 0.11  &	1.29		& 0.39	& 61.63    \\  \arrayrulecolor{gray!50}\cmidrule[0.05pt]{1-5}
				${\textit{Dataset 4}}$  & 0.93  &	16.99		& 6.9	& \hspace{0.1cm} \textminus    \\  \arrayrulecolor{gray!50}\cmidrule[0.05pt]{1-5}
				${\textit{Dataset 5}}$  & 0.79  &	10.57		& 4.12	& \hspace{0.1cm} \textminus    \\ 
				\arrayrulecolor{black}\bottomrule
			\end{tabular*}
			\label{comp_hours}
		\end{table}

		\noindent low computational times. Especially for LDA($\Sigma_{\text{KP}}$) computational time hugely increases with larger sample size and/or higher number of measurement occasions. In comparison, computational time of LDA(GEE) seems to be less affected by larger sample sizes but rather higher dimensionality (number of time points and variables). Computation of SVM results are most time-consuming, and the algorithm does not always converge after 100 iterations (Table S 6).\\ 
		Figures \ref{sim_pa} and \ref{sim_yi} show the estimates' distribution of predictive accuracy and the Youden index in the simulated data, respectively. Plots for sensitivity and specificity are shown in \\Figures S 3 and S 4, respectively. Mean (standard error) of the performance measures are also shown in Tables S 5a - e for datasets 1 - 5. 
		A first finding from Figures \ref{sim_pa} and \ref{sim_yi} is that deviation from normality (in the multivariate lognormally distributed data) in some cases increases  (dataset 1), decreases (dataset 2 and 5) the algorithms' predictive performance and Youden index, and in some cases does not have a considerable effect (dataset 3 and 4). It seems that for the scenarios with smaller sample sizes ($n_0 = 42, n_1 = 142$), no negative effect could be determined, whereas for the scenarios with much larger sample sizes ($n_0 = 948, n_1 = 1682$ and $n_0 = n_1 = 948$) there is a clear decrease in predictive accuracy and Youden index. The effect is approximately the same for all three repeated-measures LDA methods.
		A second finding is that predictive accuracy and Youden index for the SVM are visibly worse compared to the LDA methods for these imbalanced sample sizes. It has a sensitivity close to 1, but specificity close to 0, and thus mostly predicts the majority class.\\
		With respect to predictive accuracy, LDA($\Sigma_{\text{pooled}}$) without prior trimming usually performs best. Only for dataset 5 (balanced class sizes) LDA(GEE) with prior trimming (MCD algorithm) has a marginally better predictive performance in the lognormally distributed data. Values of both measures, predictive performance and Youden index, of LDA(GEE) are only equal to the other two LDA methods for dataset 5 (equal sample sizes), dataset 4 (where all methods perform poorly) and lognormally distributed data simulated based on dataset 2 (where all methods perform poorly). For dataset 1 (unbalanced classes, same temporal trends of group means), the Youden index of LDA(GEE) is higher than the values for  LDA($\Sigma_{\text{pooled}}$)  and LDA($\Sigma_{\text{KP}}$) for multivariate normally and truncated normally distributed data, especially when no trimming is applied to the training data. The boxes only slightly overlap or do not overlap at all. The reason is its higher specificity (prediction of the minority class), but its sensitivity is comparably lower.  For the lognormally distributed data generated based on dataset 1, the Youden index of LDA($\Sigma_{\text{KP}}$), especially  without prior trimming, is higher compared to the other methods, which is also due to higher specificity.\\
		It is not clear in which situations among the presented simulation scenarios trimming for outlier

		\newgeometry{bottom=15mm, top = 20mm}
		\begin{figure}[htb!]
			\begin{minipage}{0.975\textwidth} 	
				
				\hspace{0.6cm}
				\begin{minipage}{0.16\textwidth}							
					\framebox{\colorbox[RGB]{245,245,245}{\parbox{3.3cm}{ \fontfamily{qcr}\selectfont multivariate\\ normal\\ distribution}}}
				\end{minipage} 					
				\hspace{1.41cm} 	
				\begin{minipage}{0.3\textwidth}							
					\framebox{\colorbox[RGB]{245,245,245}{\parbox{3.97cm}{ \fontfamily{qcr}\selectfont multivariate lognormal distribution}}}
				\end{minipage}
				\hspace{0.075cm} 
				\begin{minipage}{0.3\textwidth}							
					\framebox{\colorbox[RGB]{245,245,245}{\parbox{3.8cm}{ \fontfamily{qcr}\selectfont multivariate\\ truncated normal distribution} }}
				\end{minipage} \\

				\begin{subfigure}{\textwidth}
					\includegraphics[scale = 0.18,trim={0cm, 0, 0cm, 0},clip]{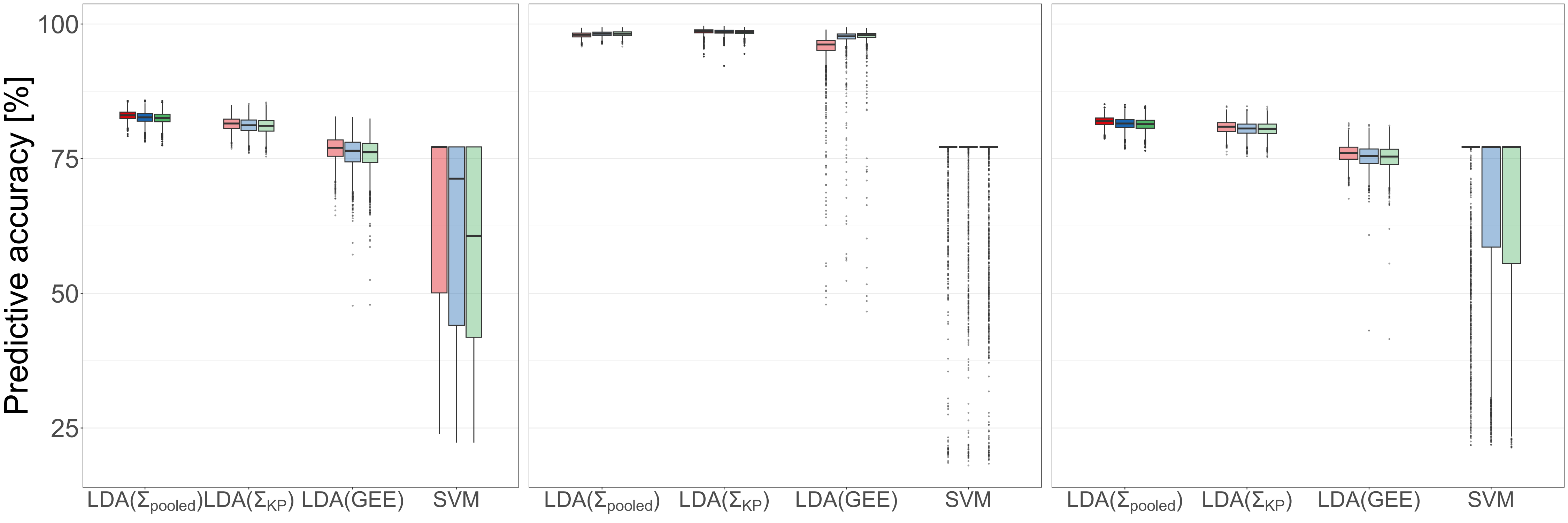}   
					\vspace{-0.25cm}
					\caption*{(a)}
					\vspace{-0.25cm}
				\end{subfigure}		
				\begin{subfigure}{\textwidth}	
					\includegraphics[scale = 0.18,trim={0cm, 0, 0cm, 0},clip]{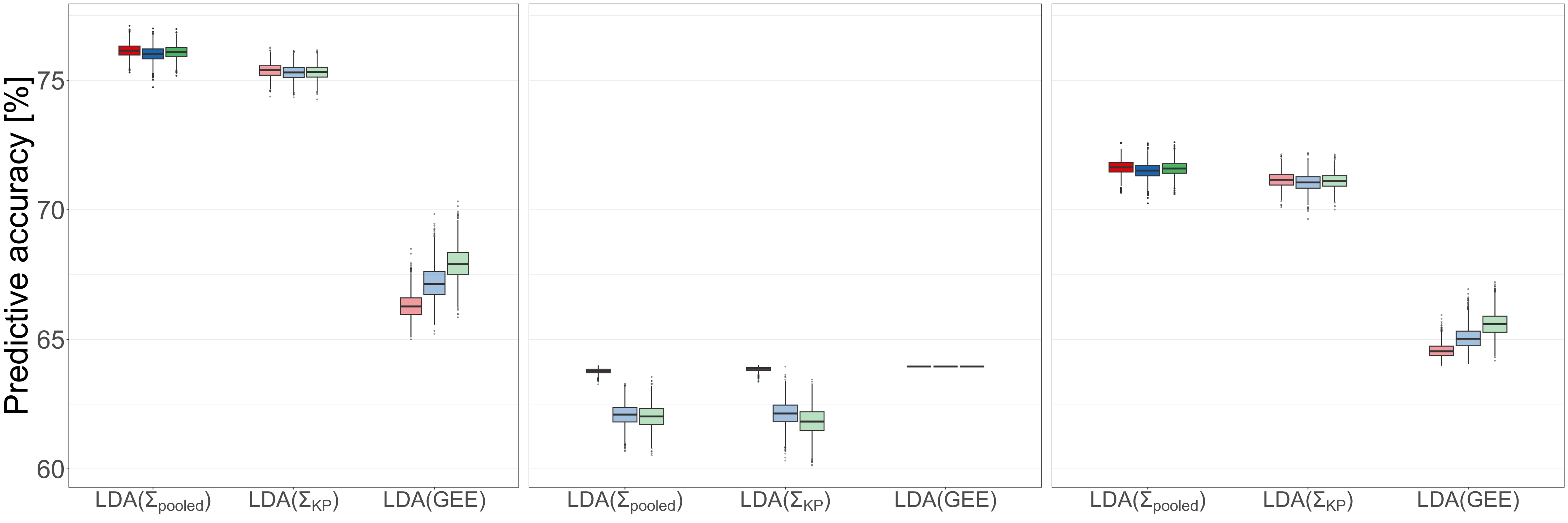} 
					\vspace{-0.25cm}
					\caption*{(b)}
					\vspace{-0.25cm}
				\end{subfigure}	
				\begin{subfigure}{\textwidth}	
					\captionsetup{labelformat=empty}	
					\renewcommand\thefigure{}
					\renewcommand{\figurename}{}
					\includegraphics[scale = 0.18,trim={0cm, 0, 0cm, 0},clip]{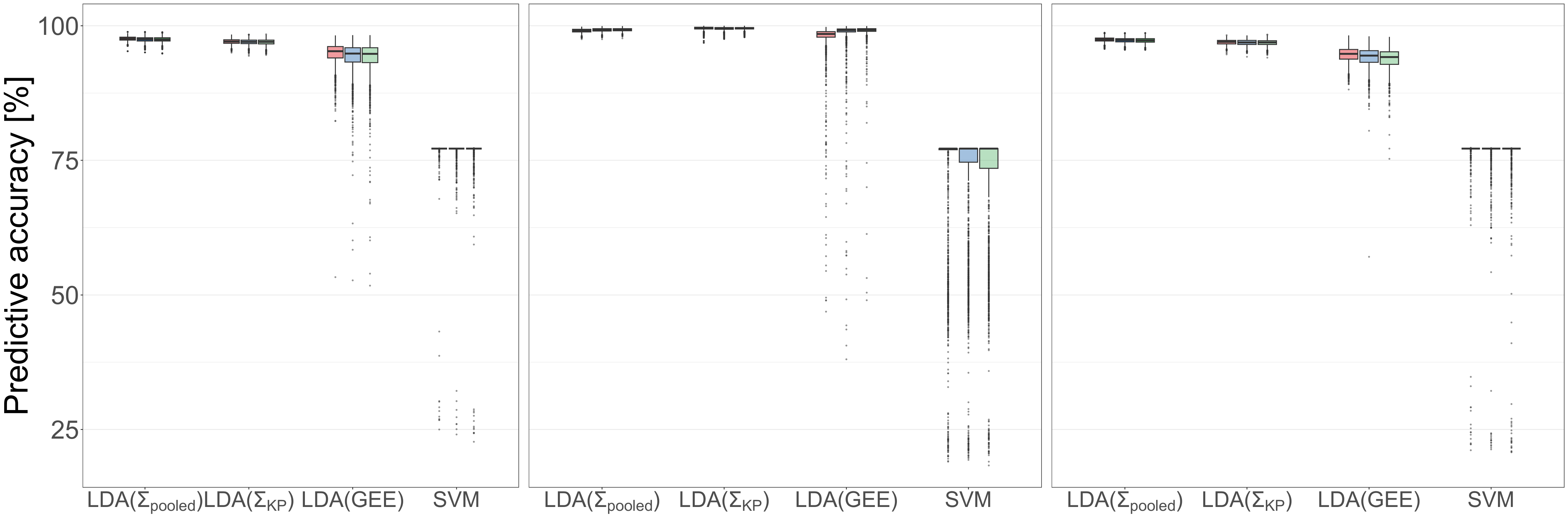} 
					\vspace{-0.25cm}
					\caption[]{(c)}
					\vspace{-0.25cm}
				\end{subfigure}	
				\begin{subfigure}{\textwidth}	
					\includegraphics[scale = 0.18,trim={0cm, 0, 0cm, 0},clip]{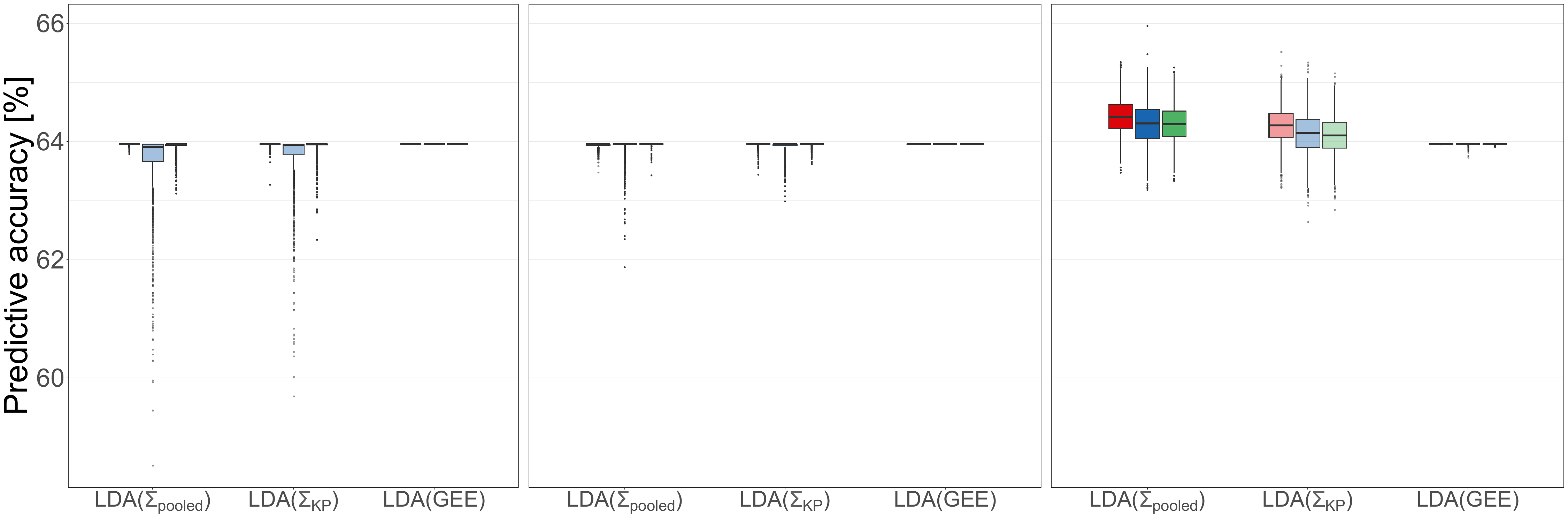} 
					\vspace{-0.25cm}
					\caption*{(d)}
					\vspace{-0.25cm}
				\end{subfigure}	
				\begin{subfigure}{\textwidth}	
					\includegraphics[scale = 0.18,trim={0cm, 0, 0cm, 0},clip]{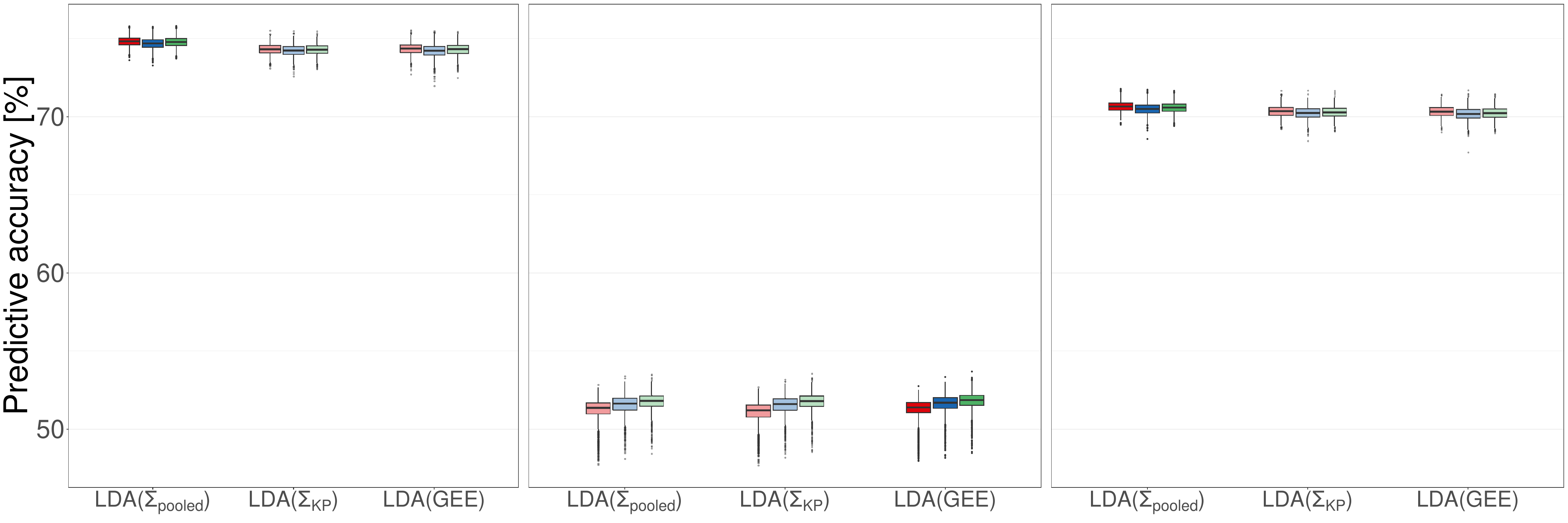} 
					\vspace{-0.25cm}
					\caption*{(e)}
					\vspace{-0.25cm}
				\end{subfigure}	
			\end{minipage}		
			\begin{minipage}{0.005\textwidth} 	
				\raggedleft
				\hfill
				\includegraphics[scale = 0.8,trim={0cm, 0, 0cm, 0},clip]{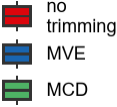}
			\end{minipage}				
			\caption*{}
		\end{figure}
		
		\restoregeometry
		
		\begin{figure}[htb!]
			\renewcommand\thefigure{5}	
			\caption[Boxplots showing the distribution of predictive accuracy per algorithm estimated in the 2000 simulated datasets.]{\fontsize{9}{12}\selectfont{(previous page) Boxplots showing the distribution of predictive accuracy estimated in the 2000 simulated datasets for the multivariate normal (left), multivariate lognormal (center) and multivariate truncated normal distribution (right). Results with the highest median value are highlighted in darker colours.\\
					(a) Dataset 1: CORE-OM dataset, group variable \textit{hospitalisation} ($n_0 = 42, n_1 = 142$)\\
					(b) Dataset 2: CASP-19 dataset, group variable \textit{loneliness} ($n_0 = 948, n_1 = 1682$) \\
					(c) Dataset 3: modified Dataset 1, such that group means collapsed over time points are equal, and group means have opposite temporal trends\\
					(d) Dataset 4: modified Dataset 2 (time points 1 \& 2), such that group means are equal, and group covariance matrices differ,\\
					(e) Dataset 5: modified Dataset 2 (time points 1 \& 2), balanced class sizes by random undersampling of group 1.}   
				{\\ \footnotesize  Abbreviations:  LDA($\Sigma_{\text{pooled}}$) - Linear discriminant analysis (pooled covariance matrix), LDA($\Sigma_{\text{KP}}$) - Linear discriminant analysis  (Kronecker product covariance matrix), LDA(GEE) - Linear discriminant analysis (covariance matrix based on generalized estimating equations estimates), SVM - Support vector machine, MVE - minimum volume ellipsoid algorithm, MCD - minimum covariance determinant algorithm.} }	
			\label{sim_pa}	
		\end{figure}

		\noindent removal may help, but there is no scenario where we explicitly simulated outliers. Both, predictive accuracy and the Youden index, somewhat increase (from a rather low performance level) for all three LDA methods in the lognormally distributed data for dataset 5 when trimming in the training data is done. }
	
	\subsection{Recommendations}
	
	 Generally, in these simulations the traditional LDA($\Sigma_{\text{pooled}}$) performs best or reasonably well with respect to predictive performance and Youden index, irrespective of smaller or larger sample size, differing group size ratios, number of measurement occasions, similar or opposite temporal trends in group means. None of the LDA methods works well for identical group means but heterogeneous covariance matrices, where they predominantly assign new observations to the majority class, and the Youden index is close to zero. The same is the case for multivariate lognormally distributed data when sample sizes are large, i.e. for an extremely evident violation of multivariate normality corresponding to extremely high values of the Mardia measure of multivariate skewness test statistic (approximately above 100).  \\
		We did not explicitly generate outliers from a different distribution than the actual data, but there may have been some random outliers. In this case, trimming for outlier removal had no effect except a minor effect on the Youden index for all LDA methods in the scenario with balanced group sizes and same temporal trends per group when data were generated from lognormally distributed data. In this case, the LDA methods still did not perform reasonably well. Multivariate trimming in the training data can be tried as a sensitivity analysis if the presence of outliers is suspected. Especially the MCD algorithm has already been recommended in the literature.\\
		In our simulations no Kronecker product covariance matrices and group means are assumed in the reference data. We used unstructured estimates of the pooled covariance matrix and group means. In our simulations, there is only an advantage of the alternative LDA($\Sigma_{\text{KP}}$) and LDA(GEE) with respect to the Youden index for data with imbalanced class sizes and comparably smaller (but not small) sample sizes. The advantage of these methods, even if no	
		underlying Kronecker product structure of the parameters can be assumed, may become more evident for smaller sample sizes. They may provide more exact estimates due to their parsimonious number of values that have to be estimated.\\ 
		Application of repeated-measures techniques should be preferred in order to incorporate the additional information about temporal trends and in order to obtain more reliable results by including data of multiple time points in the analysis provided that  moderate correlations between data of different variables and times points exist. Multicollinearity among time points and/or variables would  require removal of respective time points or variables, respectively. In case of independence between time points/variables, univariate techniques can be used.\\
		According to the psychometric literature, multivariate data are very common. An example are

		\newgeometry{bottom=15mm, top = 20mm}
		\begin{figure}[htb]
			\begin{minipage}{0.975\textwidth}

		\hspace{0.6cm}
		\begin{minipage}{0.16\textwidth}							
			\framebox{\colorbox[RGB]{245,245,245}{\parbox{3.3cm}{ \fontfamily{qcr}\selectfont multivariate\\ normal\\ distribution}}}
		\end{minipage} 					
		\hspace{1.41cm} 	
		\begin{minipage}{0.3\textwidth}							
			\framebox{\colorbox[RGB]{245,245,245}{\parbox{3.97cm}{ \fontfamily{qcr}\selectfont multivariate lognormal distribution}}}
		\end{minipage}
		 \hspace{0.075cm} 
		\begin{minipage}{0.3\textwidth}							
					\framebox{\colorbox[RGB]{245,245,245}{\parbox{3.8cm}{ \fontfamily{qcr}\selectfont multivariate\\ truncated normal distribution} }}
		\end{minipage} \\ 				
				\begin{subfigure}{\textwidth}
					\includegraphics[scale = 0.18,trim={0cm, 0, 0cm, 0},clip]{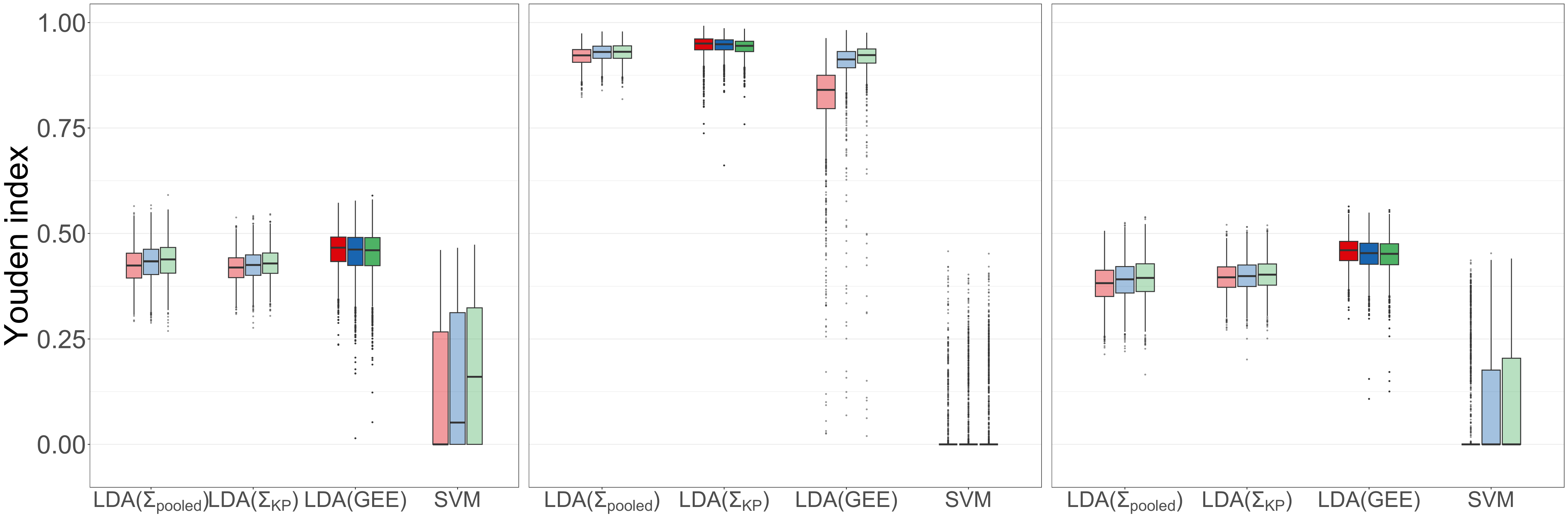}   
					\vspace{-0.25cm}
					\caption*{(a)}
					\vspace{-0.25cm}
				\end{subfigure}		
				\begin{subfigure}{\textwidth}	
					\includegraphics[scale = 0.18,trim={0cm, 0, 0cm, 0},clip]{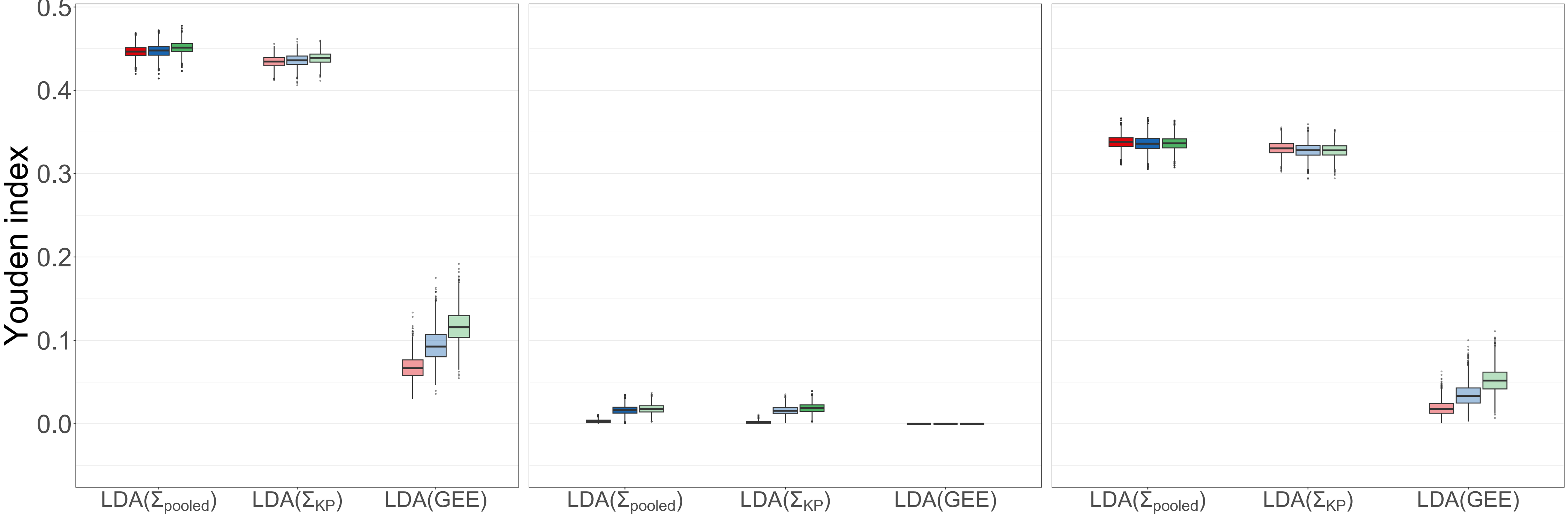} 
					\vspace{-0.25cm}
					\caption*{(b)}
					\vspace{-0.25cm}
				\end{subfigure}	
				\begin{subfigure}{\textwidth}	
					\captionsetup{labelformat=empty}	
					\renewcommand\thefigure{}
					\renewcommand{\figurename}{}
					\includegraphics[scale = 0.18,trim={0cm, 0, 0cm, 0},clip]{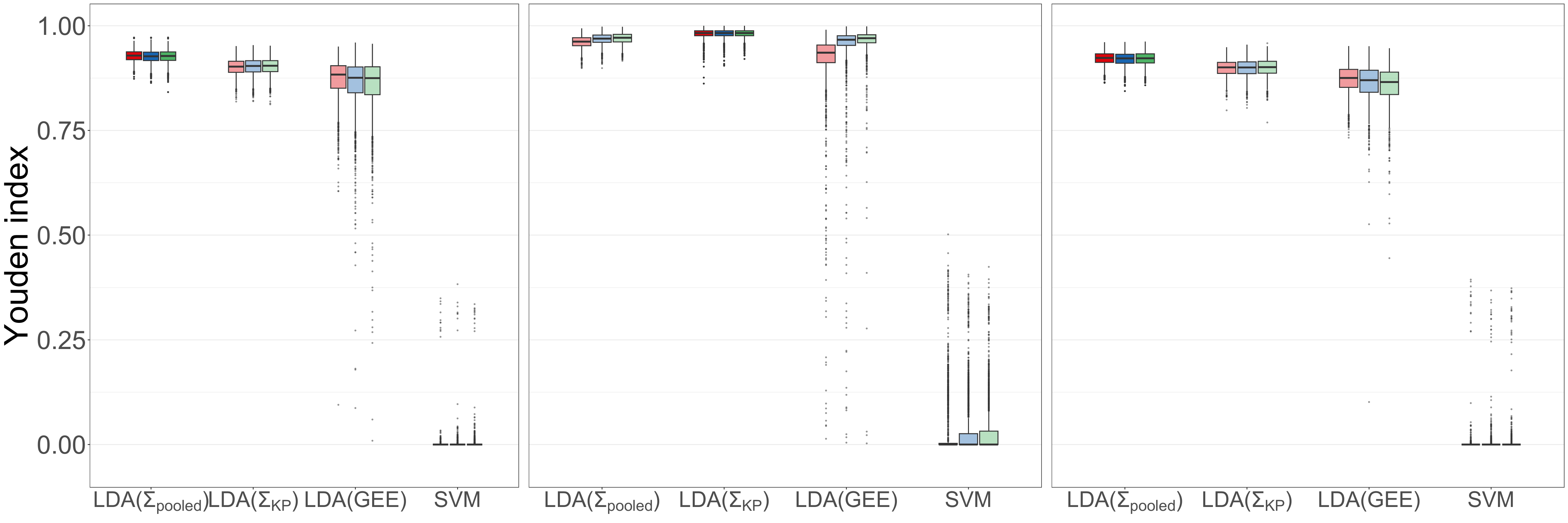} 
					\vspace{-0.25cm}
					\caption*{(c)}
					\vspace{-0.25cm}
				\end{subfigure}	
				\begin{subfigure}{\textwidth}	
					\includegraphics[scale = 0.18,trim={0cm, 0, 0cm, 0},clip]{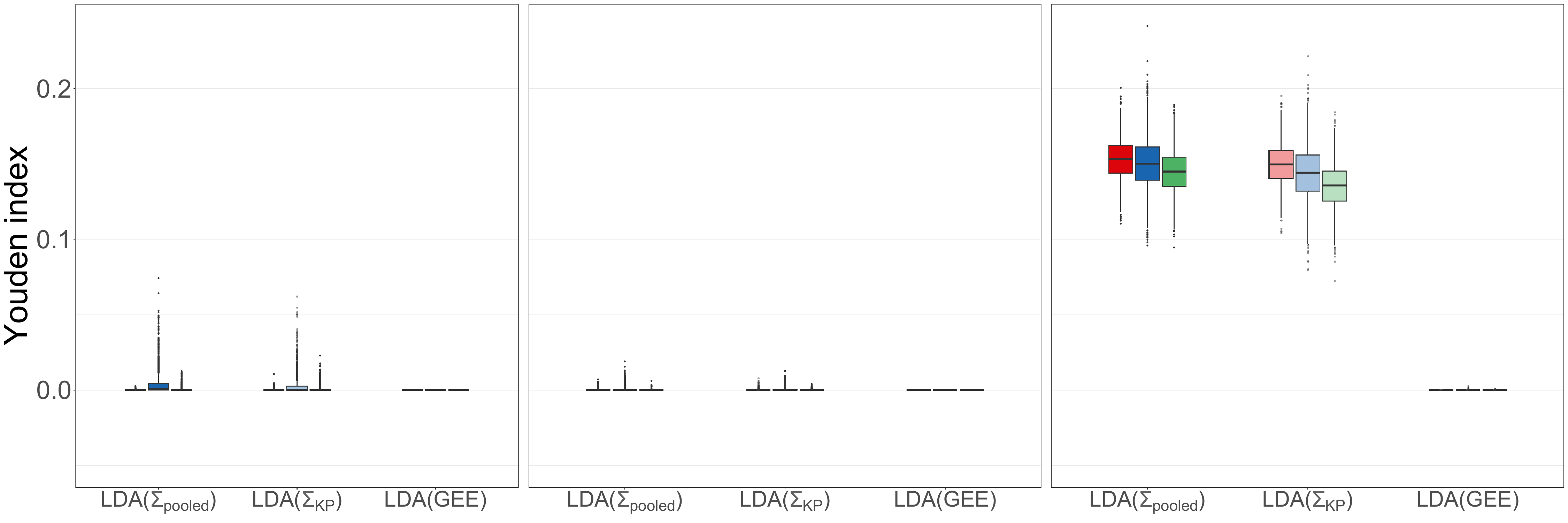} 
					\vspace{-0.25cm}
					\caption*{(d)}
					\vspace{-0.25cm}
				\end{subfigure}	
				\begin{subfigure}{\textwidth}	
					\includegraphics[scale = 0.18,trim={0cm, 0, 0cm, 0},clip]{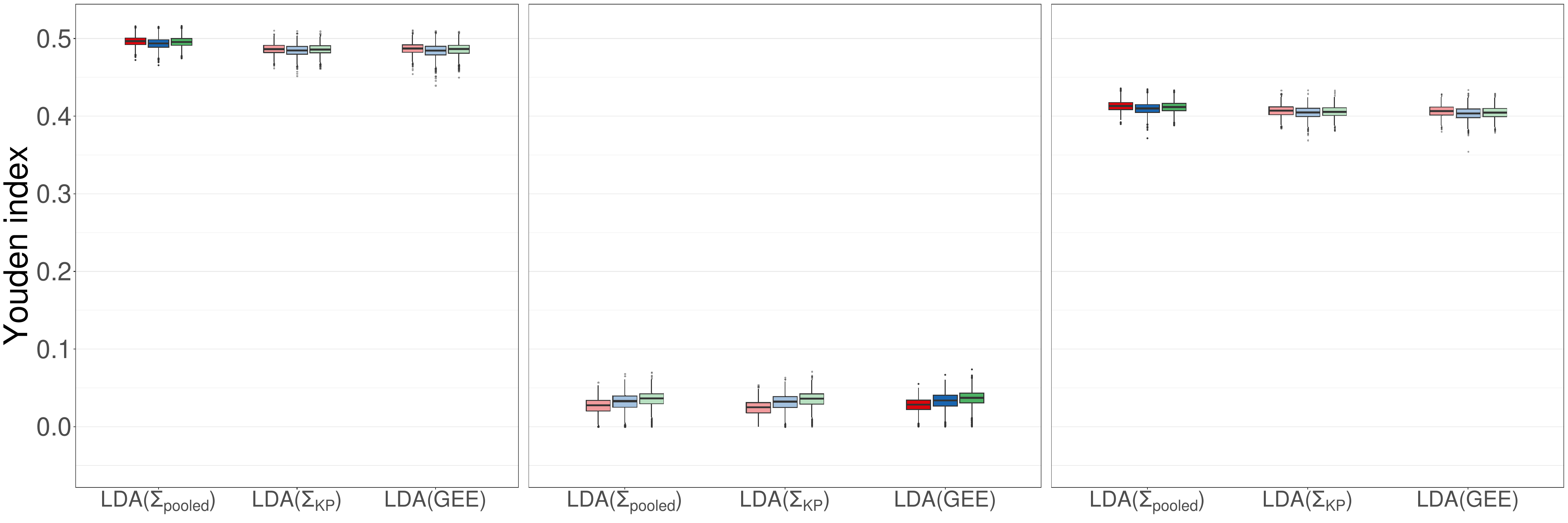} 
					\vspace{-0.25cm}
					\caption*{(e)}
					\vspace{-0.25cm}
				\end{subfigure}	
			\end{minipage}		
			\begin{minipage}{0.005\textwidth} 	
				\raggedleft
				\hfill
				\includegraphics[scale = 0.8,trim={0cm, 0, 0cm, 0},clip]{Fig5_6_legend_boxplots_simulated_data.pdf}
			\end{minipage}		
			
			\caption*{}
		\end{figure}
		
		\restoregeometry
		
		\begin{figure}[htb]
			\renewcommand\thefigure{6}	
			\caption[Boxplots showing the distribution of the Youden index per algorithm estimated in the 2000 simulated datasets.]{\fontsize{9}{12}\selectfont{(previous page) Boxplots showing the distribution of Youden index estimated in the 2000 simulated datasets for the multivariate normal (left), multivariate lognormal (center) and multivariate truncated normal distribution (right). Results with the highest median value are highlighted in darker colours.\\
					(a) Dataset 1: CORE-OM dataset, group variable \textit{hospitalisation} ($n_0 = 42, n_1 = 142$)\\
					(b) Dataset 2: CASP-19 dataset, group variable \textit{loneliness} ($n_0 = 948, n_1 = 1682$) \\
					(c) Dataset 3: modified Dataset 1, such that group means collapsed over time points are equal, and group means have opposite temporal trends\\
					(d) Dataset 4: modified Dataset 2 (time points 1 \& 2), such that group means are equal, and group covariance matrices differ,\\
					(e) Dataset 5: modified Dataset 2 (time points 1 \& 2), balanced class sizes by random undersampling of group 1.}   
				{\\ \footnotesize  Abbreviations:  LDA($\Sigma_{\text{pooled}}$) - Linear discriminant analysis (pooled covariance matrix), LDA($\Sigma_{\text{KP}}$) - Linear discriminant analysis  (Kronecker product covariance matrix), LDA(GEE) - Linear discriminant analysis (covariance matrix based on generalized estimating equations estimates), SVM - Support vector machine, MVE - minimum volume ellipsoid algorithm, MCD - minimum covariance determinant algorithm.} }	
			\label{sim_yi}		
		\end{figure}

		\noindent  the widely applied  questionnaires using Likert-type responses where multiple correlated aspects related to  
		an overall topic are measured. In order to assess the usefulness of different sets of variables for distinguishing two classes of individuals, LDA can be applied for class prediction and its performance for different sets of variables can subsequently be compared to determine the most relevant variables. Usually, for LDA applied to cross-sectional data, Fisher  discriminant function coefficients \citep{Fisher1936} are computed in order to assess relative variable importance within a particular set. The method can in principle also be applied to repeated measures data. It does not assume multivariate normality although it requires homogeneity of covariance matrices.   
	
	\section{Conclusions}\label{conclusion}
	
	Longitudinal studies are conducted in psychology and other disciplines. Data in psychology and the social sciences are often characterized by nonnormal distributions, especially skewness. LDA is widely applied as a standard technique in these fields, e.g. to questionnaire data where answers are measured on Likert scales that are summarized in subscales based on means or sums of multiple Likert items (i.e. single questions), either for classification tasks or for identifying variables most relevant to group separation. Repeated measures techniques are preferable for the analysis of data that are collected repeatedly over time compared to conducting several independent analyses for each time point in case temporal correlations exist.\\
	We compared the performance of robust repeated measures DA techniques proposed by Brobbey et al. (\citeyear{Brobbey2021, Brobbey2022}) and the longitudinal SVM by Chen 
	and Bowman (\citeyear{Chen2011}) using multiple performance measures.  We based these comparisons on real psychometric datasets which differ with respect to sample size, sample size ratio, class overlap, temporal variation, number of repeated measurement occasions, and properties of group means and covariance matrices. We thus considered additional scenarios to those in Brobbey et al. (\citeyear{Brobbey2021, Brobbey2022}), where Kronecker product structures of means and covariances and thus constant correlations and means of the variables over time were assumed. We also compared several alternative methods among each other in contrast to comparing a particular alternative to the standard method at a time. We included the longitudinal SVM because it is similar to repeated measures LDA in that they are both linear classifiers for which variable weights can additionally be computed and temporal correlations are considered in the analysis. We did not consider extensions of other supervised machine learning algorithms for classification since they usually assume independence between time points \citep{Ribeiro2019} and do not have a comparably intuitive interpretation of variable weights as the linear SVM. \\
	We followed the guidelines for neutral comparison studies by Weber et al. (\citeyear{Weber2019}) and the general design of simulation studies by Morris et al. (\citeyear{Morris2019}). We found that the alternative robust methods may not be required for sufficiently large sample sizes and  absence of outliers.  Limitations of our simulation study are that only a limited number of scenarios and datasets are considered.  Further examination in data with smaller sample sizes and in data containing outliers from a different distribution  would be helpful. In this context, the influence of different choices for the trimming parameter when applying one of the trimming algorithms for outlier removal may also be examined.  To date, no recommendations on the choice of the trimming parameter for multivariate data exist. Therefore, for an actual dataset, multiple values should be tried.
	 Moreover, due to availability of suitable datasets in particular  given data protection policies, and limited number of scenarios considered in every simulation study in general, further conclusions may be possible when applying the methods to other datasets.  As with any simulation study, our results can therefore not be generalized beyond the considered scenarios.
	We found that none of the LDA methods did work well for extreme deviations from normality, and heterogeneity of covariance matrices when group means were identical, respectively. Conclusions based on the performance in the reference datasets and based on data simulations, respectively, are similar.


	%
	%
	%
	%

	%
	%
	
	%
	%

	\vfill
	\section*{Declarations}
	
	
	\begin{itemize}
		\item[] Acknowledgments: The authors gratefully acknowledge the resources on the LiCCA HPC cluster of the University of Augsburg, co-funded by the Deutsche Forschungsgemeinschaft (DFG, German Research Foundation) \textminus Project-ID 499211671.
		\item[] Funding: No funding was received to assist with the preparation of this manuscript.
		\item[] Conflict of interest: The authors have no competing interests to declare that are relevant to the content of this article.
		\item[] Ethics approval: Not applicable. 
		\item[] Code availability:  Supplementary files containing the \texttt{R} code used for data simulations can be found on Figshare (\url{https://figshare.com/s/104aeb2a870a810f80bd}).
	\end{itemize}

	
	\bigskip

		\clearpage
	\bibliographystyle{apa}
	\bibliography{QCMB_article}
	

\end{document}